\documentclass[12pt]{article}
\usepackage{amsmath,amssymb}
\usepackage{type1cm}
\usepackage{amsthm}
\usepackage[dvipdfmx]{graphicx}
\RequirePackage[l2tabu, orthodox]{nag} 
\usepackage[all, warning]{onlyamsmath}
\usepackage[subrefformat=parens]{subcaption}
\usepackage{caption}
\usepackage{bm}
\usepackage{authblk}
\usepackage{url}
\usepackage{color}
\usepackage{amsfonts}
\usepackage{here}
\newcommand{\argmax}{\mathop{\rm argmax}\limits}
\newcommand{\argmin}{\mathop{\rm argmin}\limits}

\newcommand{\indep}{\mathop{\perp\!\!\!\perp}}
\usepackage{amsthm}
\theoremstyle{remark}
\newtheorem{thm}{Theorem}
\newtheorem{assumption}{Assumption}
\usepackage{subcaption}
\title{Regression Discontinuity Design with Multiple Groups for Heterogeneous Causal Effect Estimation}

\author{Takayuki Toda\thanks{Graduate school of Economics, Keio University} \and Ayako Wakano\thanks{ Department of Economics, Tokai University} \and Takahiro Hoshino\thanks{ Department of Economics, Keio University / RIKEN AIP center} }
\date{\today}

\begin{document}
\maketitle
\thispagestyle{empty}
\begin{abstract}
We propose a new estimation method for heterogeneous causal effects which utilizes a regression discontinuity (RD) design for multiple datasets with different thresholds. The standard RD design is frequently used in applied researches, but  the result is very limited in that the average treatment effects is estimable only at the threshold on the running variable.
In application studies it is often the case that thresholds are different among databases from different regions or firms. For example thresholds for scholarship differ with states.
The proposed estimator based on the augmented inverse probability weighted local linear estimator can estimate the average effects at an arbitrary point on the running variable between the thresholds under mild conditions, while the method adjust for the difference of the distributions of covariates among datasets.
We perform simulations to investigate the performance of the proposed estimator in the finite samples.\\
\begin{enumerate}
\renewcommand{\labelenumi}{{\it Keywords:}}
\setlength{\leftskip}{1cm}
 \item Regression discontinuity design; Heterogeneous causal effects; Counterfactual; Double robustness; Nonparametric regression;
\end{enumerate}

\end{abstract}

%%\newpage
%%\tableofcontents
%%\newpage

\newpage

\pagestyle{plain}
\setcounter{page}{1}

\section{Introduction}
Regression discontinuity (RD) design originated in Thistlethwaite and Campbell (1960) that study the effect of the student scholarships on future academic outcomes. In RD design, for evaluation of the  intervention of which status is determined by whether a covariate exceed a fixed known threshold or not, subjects with values just below the threshold and those above the threshold are compared,  where the intervention status is as good as randomly assigned.

RD design is well applied by empirical researchers to estimate the treatment effect at the target population, similar to other quasi-experimental methods. Applications of RD design are found in various empirical fields in economics such as labor, public, education, and development economics. Detailed literature survey is found in Imbens and Lemieux (2008), and in Lee and Lemieux (2010).

As there have been numerous empirical applications of RD design, more methodological and theoretical extensions are suggested in different directions such as the case of fuzzy discontinuity by Hahn et al. (2001), as for the selection of bandwidth (Ludwig and Miller, 2007; Imbens and Kalyanaraman, 2012; Calonico et al., 2014; Arai and Ichimura, 2018), and for different tests for estimations (Lee, 2008). 

Our goal in this paper is to propose a new method for estimation of counterfactual functions and heterogeneous causal effects considering a RD design with multiple groups which have different thresholds. In the standard RD designs, one of the serious limitations is that the intervention effect only at the discontinuity point is evaluable. Angrist and Rokkanen (2015) proposed a method for identification of the causal effects away from the cutoff. However, their approach is that the running variable is assumed to be ignorable if conditional on the other available predictors, and it is different from our attempt to estimate the counterfactual functions themselves. To consider what kind of assumption and estimation method are needed is an important task in this research. 
In addition, we provide a method for optimization of threshold as an application of our method. Most past studies have considered thresholds as a fixed value and not dealt with threshold itself as an object of study. However, the real interest of researchers should lie not only in evaluation of past interventions under a given threshold but also in an appropriate threshold setting as a support for decision-making for future interventions. Therefore, we develop a method to estimate an optimal threshold in terms of cost effectiveness.

Our methodological development is closely related to the multiple thresholds in RD method.
Indeed, the empirical literature utilizes the standard RD design with single threshold. However, it is not uncommon to have multiple thresholds in actual datasets. We often observe multiple thresholds to assign one treatment in a target population. For example, it is often the case that local governments determine the cutoff value for running variables such as test scores, poverty indexes, birth weight, geolocation, and income. When the different administrative districts set each unique threshold of admission test score, it leads multiple thresholds exist in the target population (Lucas and Mbiti, 2014). Similarly, the geographical division often sets own eligible cutoff value for social welfare programs (Crost et al., 2014). 
In Japan, age limits of the local goverments' programs to make medical expenses for children free vary by the local governments.
In this way, countless situations are applicable for multiple thresholds, while RD application is merely concentrated on single threshold method.

There is scarce methodological literature that handles multiple threshold situations. The past literature that deal with multiple thresholds is Papay et al. (2011). Papay et al. (2011) shows how to incorporate multiple dimensions of running variables in the RD design with single dataset, which is different from our model setup with multiple datasets. Literature has also moved to the situations where thresholds or cutoff points were unknown for researchers (Henderson et al., 2014; Porter and Yu, 2015; Chiou et al., 2018). Our method is clearly differed from their works as we assume the situation where value of cutoff is observed from datasets. 

This paper is organized as follows.
In Section 2 we describe the standard RD design settings as basics of the proposed method. In addition we provides the details of our design that using a special structure that there are multiple groups with different thresholds makes it possible to estimate counterfactuals and causal effects.
Section 3 we propose a new AIPW kernel estimator making the best use of the observed data in our design. 
In Section 4 we investigate the asymptotic properties of the estimator proposed in Section 4 and show its double-robustness. 
In Section 5 we provide a method to estimate an optimal bandwidth as an application of our method.
In Section 6 we report a simulation for studying the properties of the proposed estimator in the finite sample.
In Section 7 we summarize this paper and discuss future outlook on this research.

\section{Model}
In this section, we briefly summarize framework and theory of the conventional regression discontinuity design. Then we extend the discussion to the case with multiple thresholds and propose our method to estimate the unobservable counterfactuals in the conventional RD designs and heterogeneous causal effects by using them. In this paper consider only the situation where there is just two groups for simplicity. Our discussion and notation are based on Imbens and Lemieux (2008) and modern literature using Rubin Causal Model (RCM) setup with a concept of potential outcomes (Rubin, 1974; Holland, 1986).

\subsection{Regression discontinuity design\label{sec:}}
As is the usual case with RCM, consider the situation that there are two types of interventions, special intervention (i.e. treatment) and normal intervention (i.e. control), and researchers are interested in the causal effect of the intervention. Corresponding to those two types of interventions, there are two potential outcomes for each unit. Denote by $Y_{ji}$ the potential outcomes of unit $i\in N$, where $N=\{1,...,n\}$ is a set of $n$ units, and the potential outcome for treatment is $Y_{1i}$ and the potential outcome for control is $Y_{0i}$.

Now let the intervention assignment indicator of unit $i$ denote $Z_i\in\{0,1\}$, which is 1 when unit is exposed to treatment and 0 otherwise. The observed outcome variable can be expressed as
\begin{eqnarray}
  Y_i & = & Z_i Y_{1i}+(1-Z_i) Y_{0i}=
  \begin{cases}
  Y_{1i}~~~if~Z_i=1\\
  Y_{0i}~~~if~Z_i=0
  \end{cases} (i=1,...,n).
\end{eqnarray}
In addition, let a finite dimensional vector of pretreatment covariate variables except $X_i$ denote $\bm{W_i} \in \mathbb{R}^m$.

In the setting of RD designs, the type of intervention allocated to unit $i$ is determined by whether a running variable $X$ is above a threshold $c$. 
RD designs are generally divided into two types, the sharp RD (SRD) design and the fuzzy RD designs depending on how to determine the assignment of intervention. In this study we limit the discssion to the sharp RD design. In the sharp RD design the assignment $Z_i$ is based on a deterministic function of the running variable $X_i$ defined as 
\begin{eqnarray}
Z_i=1(X_i>c).
\label{eq:z_func}
\end{eqnarray}
Under this function, all the units observing $X_i$ above $c$ are exposed to treatment and the others are exposed to control. 

In the sharp RD design, although the running variable $X_i$ does not overlap between the treatment group and the control group, the assignment $Z_i$ is only depending on $X_i$, therefore Missing at random (MAR) (Rubin, 1976), that is,  
\begin{eqnarray}
Y_{1i},Y_{0i} \indep Z_i|X_i, 
\label{eq:strig}
\end{eqnarray}
is satisfied.

Under MAR, if the models of $E(Y_1|X)$ and $E(Y_0|X)$ are parametric, $E(Y_1|X)$ can be extrapolated even below the threshold and $E(Y_0|X)$ also can be extrapolated above the threshold. Therefore, $E[Y_1-Y_0|X=a]$ at any arbitrary point $X=a$ can be estimated and $E[Y_1-Y_0]$ also can be. 
However, nonparametric regression do not permit extrapolation and only $E(Y_1|X)$ for $X>c$ and $E(Y_0|X)$ for $X<c$ and the difference of those at the discontinuity point ,that is, the local average treatment effect (LATE) 
\begin{align}
\begin{split}
\tau_{SRD}&=E[Y_{1}-Y_{0}|X=c]=E[Y_{1}|X=c]-E[Y_{0}|X=c]\\
&=E[Y|X=c,Z=1]-E[Y|X=c,Z=0]
\end{split}
\end{align}
can be estimated. This is the main goal in the RD designs. However only the treatment group can include units who observe $X_i=c$ and the control group cannot, hence the conditional expectation of the observed outcomes $Y_i$ given $X_i$ is discontinuous at $c$. Thus $\tau_{SRD}$ can be regarded as
\begin{align}
\tau_{SRD}=\lim_{x\downarrow c}E[Y|X=x]-\lim_{x\uparrow c}E[Y|X=x],
\end{align}
and obtained by point estimations of the limits from the left and right.

RD design is useful in many practical cases, however it is one of the major limitations that only LATE at the discontinuity point can be estimated and thus the result may lack generalizability (Lee and Lemieux, 2010). This problem is due to the structure that there is no overlap in $X_i$ between the treatment and control groups and the counterfactual cannot be obtained. To solve this problem at least partially, we propose a new method when different datasets with different thresholds are available.

\subsection{Regression discontinuity design with two groups}
In this paper, to estimate the unobserved potential outcome in the standard RD design (i.e. counterfactual), we consider the RD designs with multiple groups which have different thresholds. We assume the case where the same intervention is provided to several groups (e.g. geographical regions) and those groups have different thresholds from each other on a same running variable and the types of intervention for units are determined by the thresholds of the groups to which they belong. Other basic settings are the same as the case with the standard RD design described in the previous part; there are two types of intervention, or treatment and control, and corresponding to those interventions there are two potential outcomes, and we focus only on the sharp RD design.

\begin{figure}[tbhp] 
  \centering
   \includegraphics[width=90mm]{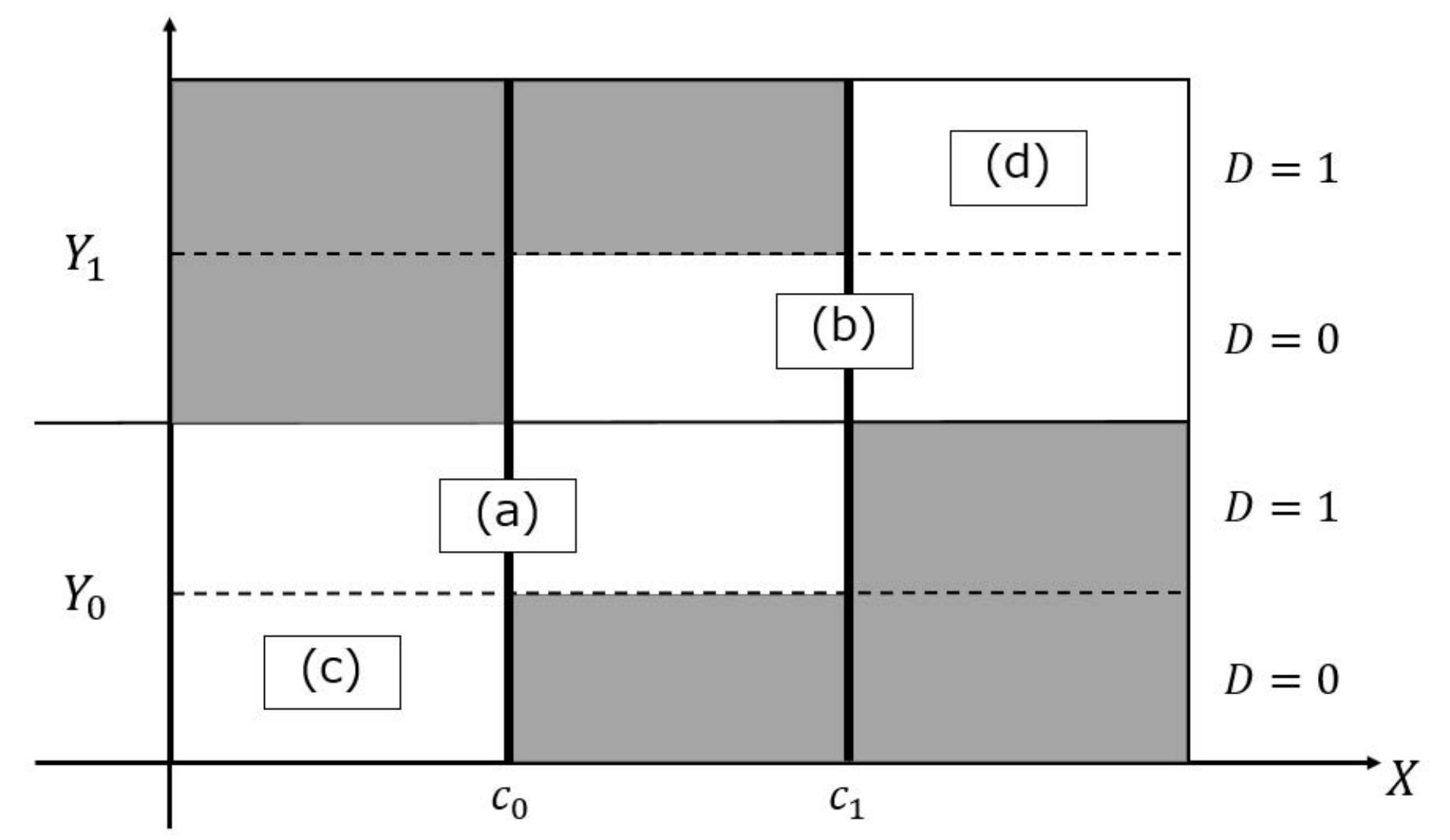}
 \caption{Observed and unobserved outcomes. The white areas show where outcomes can be observed and the gray ones show where outcomes cannot be observed. For $X<c_0$, $Y_0$ can be commonly observed in both data sets and $Y_1$ cannot. For $X>c_1$, on the contrary, only $Y_1$ can be observed and $Y_0$ cannot. For $c_0<X<c1$, $Y_1$ for $D=0$ and $Y_0$ for $D=1$ are observed and $Y_0$ for $D=0$ and $Y_1$ for $D=1$ are missing. This study utilizes this symmetric structure for  estimation of the counterfactual functions.}
 \label{fig:po}
\end{figure}

In the following, we consider only the case with two groups. Each unit belongs to either of the two groups. Let the group assignment indicator for unit $i \in N$, where $N=\{N_0,N_1\}=\{1,...,n_0,n_0+1,...,n_0+n_1\}$, $N_0=\{1,...,n_0\}$ and $N_1=\{1,...,n_1\}$, be denoted by $D_i\in\{0,1\}$, which takes 0 if $i\in N_0$ and takes 1 if $i\in N_1$. In addition, $c_k(k=0,1; c_0<c_1)$ denotes the thresholds of the two groups, $c_0$ is the one in the group of $N_0$ and $c_1$ is the other. By using the subscript $d_i \in \{0,1\}$ representing the group to which unit $ i $ belongs, the function of intervention assignment is 
\begin{eqnarray}
Z_i=1(X_i>c_{d_i}).
\end{eqnarray}
According to this function, observable outcomes for unit $i$ from $N_0$ are $Y_{0i}$ for $X_i \leq c_0$ and $Y_{1i}$ for $X_i>c_0$, and for unit $i$ from $N_1$, $Y_{0i}$ for $X_i\leq c_1$ and $Y_{1i}$ for $X_i>c_1$ are obsereved. Thus, different potential outcomes are observed depending on the groups for $c_0<X_i<c_1$, while $Y_{0i}$ for $X_i<c_0$ and $Y_{1i}$ for $X_i>c_1$ are commonly observed from both of the two groups, as shown in the Figure\ref{fig:po}.

\begin{figure}[tbhp] 
 \centering
   \includegraphics[width=0.6\linewidth
]{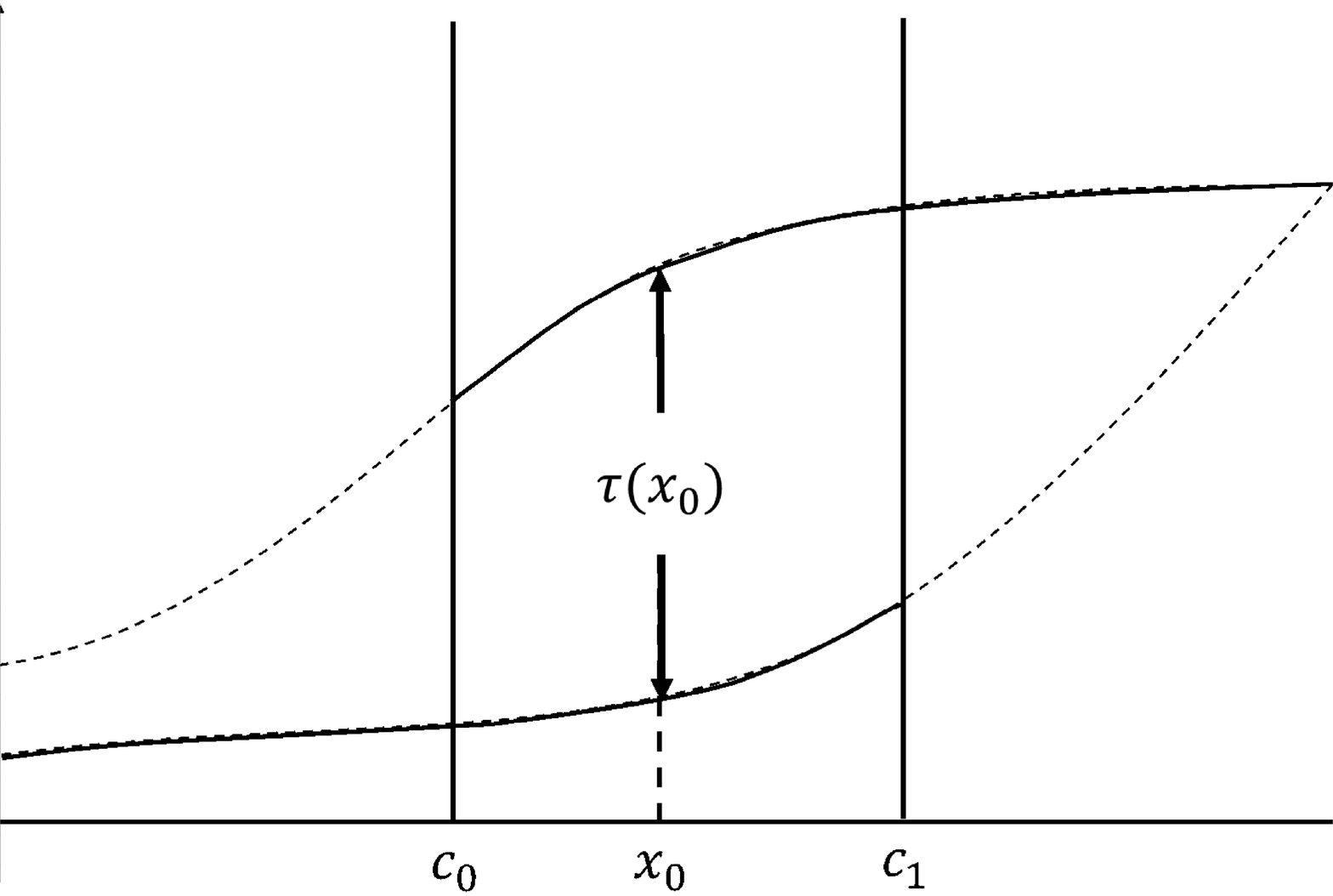}
 \caption{Causal effects at the point $X=x_0$. If we can estimate $g_1(x)$ for $X>c0$ and $g_0(x)$ for $X<c1$, the causal effect can be defied as the function of $X$ formed by $\tau(x) = g_1(x)-g_0(x)$ for $c_0<X<c_1$.}
 \label{fig:ce}
\end{figure}

Now, let the conditional expectation functions given $X_i$ depending on the group assignment be denoted by
\begin{eqnarray}
E[Y_j |X=x,D=k]=g_{jk}(x)~~~(j=0,1; k=0,1).
\label{eq:func_ind}
\end{eqnarray}
This expression allows the regression function to be different by the group assignment, however we are not interested in the individual functions for each group. Our main interest lies in the functions in the target common population: 
\begin{eqnarray}
E[Y_j |X=x]=g_j(x)~~~(j=0,1)
\label{eq:cefun}
\end{eqnarray}
Especially, between the two thresholds, both of the potential outcomes $Y_1$ and $Y_0$ are observed and thus it should be potentially possible to estimate $g_1(x)$ and $g_0(x)$ for $c_0<X<c_1$, which overlap each other. If we can estimate them, we can also estimate the average treatment effects at arbitrary points between the two thresholds defined as 
\begin{align}
\begin{split}
\tau(x) &= E[Y_1-Y_0|X=x]\\
&= E[Y_1|X=x]-E[Y_0|X=x] \\
&= E[Y|X=x,Z=1]-E[Y_0|X=x,Z=0] \\
&= g_1(x)-g_0(x) ~~~ (c_0<x<c_1),
\end{split}
\label{eq:ce}
\end{align}
as shown in the Figure\ref{fig:ce}.

\begin{figure}[tbhp]
 \begin{minipage}{0.5\hsize}
  \centering
   \includegraphics[width=\linewidth]{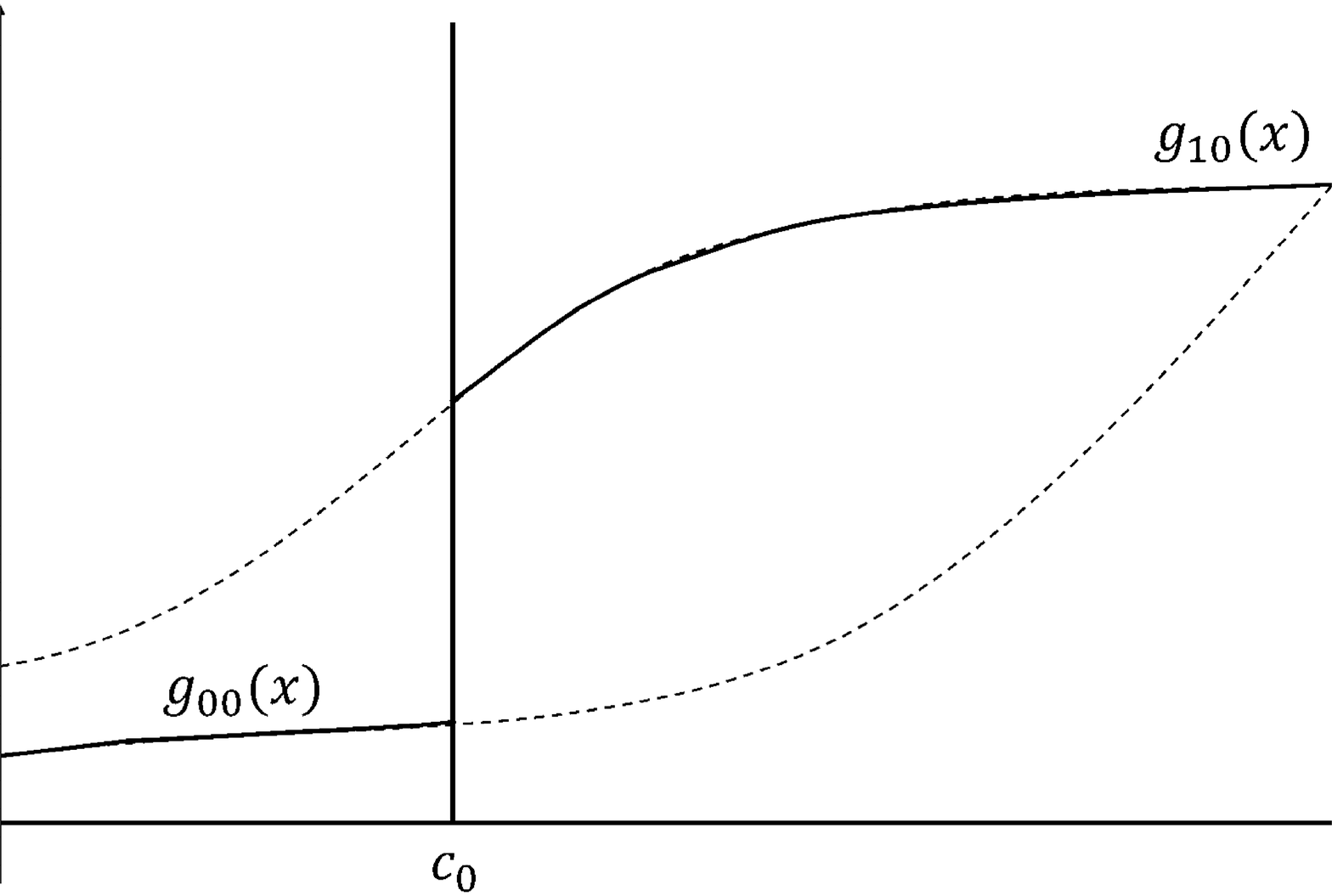}
   \subcaption{}
 \end{minipage}
 \begin{minipage}{0.5\hsize}
  \centering
   \includegraphics[width=\linewidth]{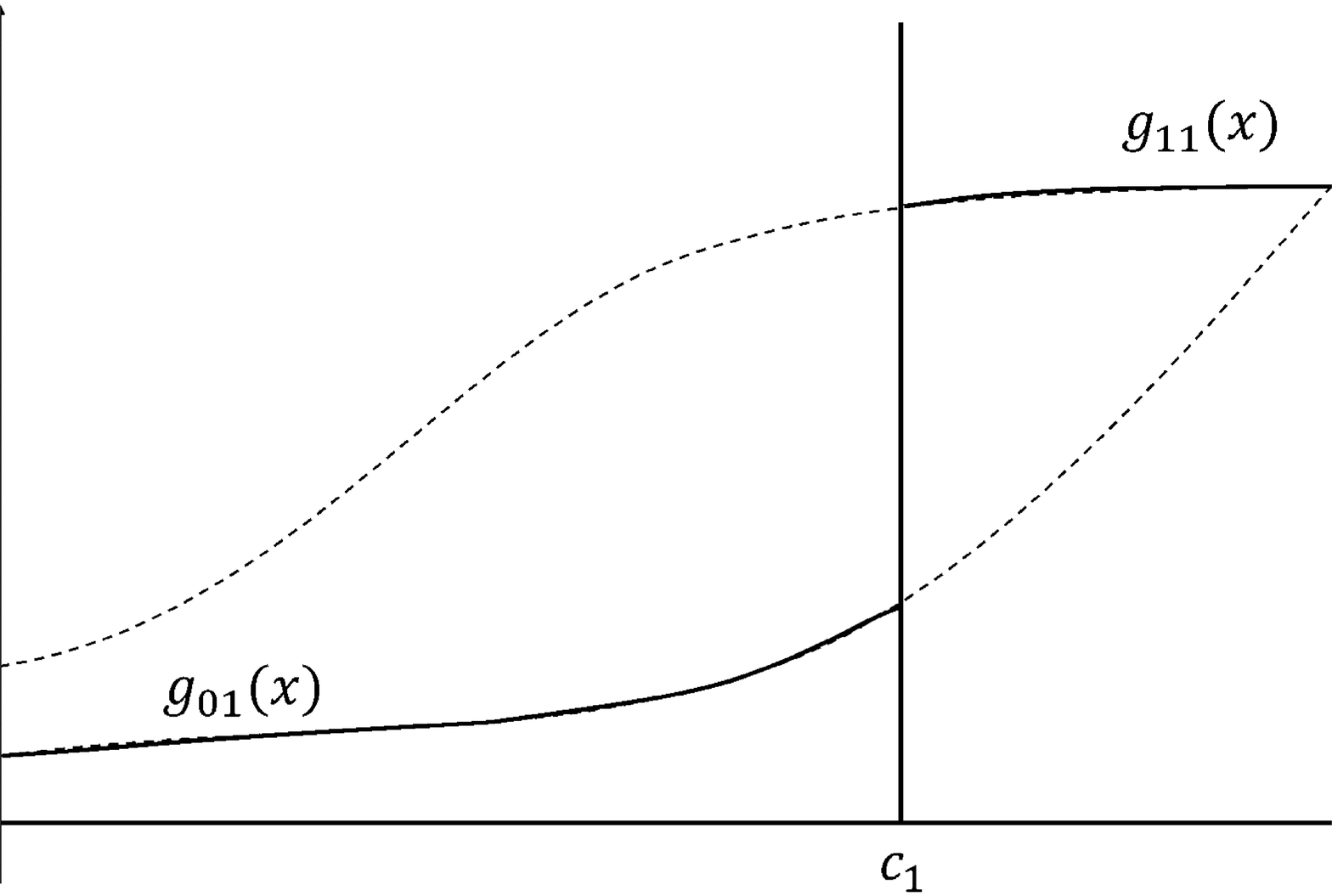}
   \subcaption{}
 \end{minipage}
 \caption{Conditional expectation functions of potential outcomes. The solid line indicates the parts that can be estimated from the data and the dashed line indicates the parts cannot be estimated. The left panel is for the case using $D=0$ and the right panel is for $D=1$.}
 \centering
 \label{fig:obs_po}
\end{figure}

Nevertheless what can be estimated from the data is only function (\ref{eq:func_ind}) and we cannot estimate function (\ref{eq:cefun}) directly. The conditional expectation (\ref{eq:cefun})  can be rewritten as
\begin{align}
\begin{split}
E[Y_1|X=x]=&E[Y_1|X=x,D=0]Pr(D=0|X=x)\\
&\qquad+E[Y_1|X=x,D=1]Pr(D=1|X=x)\\
E[Y_0|X=x]=&E[Y_0|X=x,D=0]Pr(D=0|X=x)\\
&\qquad+E[Y_0|X=x,D=1]Pr(D=1|X=x)
\end{split}
\label{eq:rece}
\end{align}
and if we knew all factors of the right hand side in equation (\ref{eq:rece}), we could estimate function (\ref{eq:cefun}) following equation (\ref{eq:rece}). However, the observed potential outcome is limited as described above, what can estimate directly from the data are only
\begin{align}
\begin{split}
E[Y_0|X=x,D=0]=g_{00}(x)~~~(x<c_0),~~~E[Y_1 |X=x,D=0]=g_{10}(x)~~~(x>c_0)\\
E[Y_0|X=x,D=1]=g_{01}(x)~~~(x<c_1),~~~E[Y_1 |X=x,D=1]=g_{11}(x)~~~(x>c_1)
\end{split}
,
\end{align}
and the other parts cannot be estimated directly in general, as shown in Figure \ref{fig:obs_po}.
Therefore, whereas $g_0(x)$ for $x<c_0$ and $g_1(x)$ for $x>c_1$ can be estimated according to the equations (\ref{eq:rece}), 
$g_{00}(x)$ and $g_{11}(x)$ for $c_0<x<c_1$ cannot be estimated
and thus we cannot estimate $g_0(x)$ and $g_1(x)$ between the thresholds of most interest.
In what follows, we consider what kind of assumption is necessary to realize unbiased estimation of the function $g_j(x)$.

First, consider the most optimistic situation, where the group is randomly assigned for units and the estimated functions are independent of the data assignment. In this case, $E[Y_j|X=x,D=k]=E[Y_j|X=x]$ holds and using only one data from observed group of either $D=0$ or $D=1$ does not generate bias. 
One of the situations in which this assumption holds is where a type of randomized controlled trial (RCT) can be conducted, where units are randomly distributed to two groups with different thresholds. However, in the field of medicine or social science  such as ecnomics, there are not many situations where it is possible to implement random assignment for structural or ethical reasons. 

In the following, we investigate the case where the conditions that group assignment is randomly determined and the conditional expectation functions do not depend on the group assignment are not satisfied; that is,  
\begin{align}
g_{jk}(x)= E[Y_j|X=x,D=k]\not=E[Y_j|X=x]=g_j(x).
\end{align}
This means that there is a selection bias between the two groups. In this case, a standard approach can cause biased estimates. In order to achieve the unbiased estimator of the conditional expectation functions, we additionally assume ignorability. 
\begin{assumption}
(Ignorability)\\
The group assignment variable $D$ depends only on the covariates $X$ and $W$ but not on the outcome variable $Y_0$ and $Y_1$.
\begin{align}
Pr(D|Y_0,Y_1,X,W)=Pr(D|X,W).
\label{eq:ign}
\end{align}
In form of the conditional independence given $X$ and $W$,
\begin{align}
Y_1,Y_0 \indep D|X,W.
\end{align}
\label{as:ign}
\end{assumption}

This assumption also can be rewritten in another way using Bayes' theorem. 
\begin{align}
Pr(Y_0,Y_1|D,X,W)=Pr(Y_0,Y_1|X,W)
\end{align}
In this form it can be interpreted as meaning that given the covariates $X$ and $W$ the simultaneous distribution of $Y_0$ and $Y_1$ is independent of the group assignment $D$.

Under this assumption, the conditional expectations satisfy
\begin{align}
\begin{split}
E(Y_0|X) &= E_{W|X}[E(Y_0|X,W)]\\
&= E_{W|X}[E(Y_0|X,W,D=1)]\\
&= E_{W|X}[E(Y_0|X,W,D=1,Z=0)]\quad(X<c_1)\\
E(Y_1|X) &= E_{W|X}[E(Y_1|X,W)]\\
&= E_{W|X}[E(Y_1|X,W,D=0)]\\
&= E_{W|X}[E(Y_1|X,W,D=0,Z=1)]\quad(X>c_0)
\end{split}
.
\end{align}
However, when the covariates $W$ is high-dimensional, as in many cases, correct identification of parametric function form is mostly impracticable, furthermore, if using nonparametric regression including the local linear kernel regression, practitioners are faced with the problem known as the Curse of Dimensionality\footnote{The Curse of Dimensionality is the phenomena that the amount of data required for estimation increases exponentially when there are many explanatory variables (Hoshino, 2009). More specifically, let $d$ denote the number of dimension, then asymptotic mean squared error is proportional to $N^{-4/(d+4)}$ (H\"{a}rdle et al., 2004).}. To avoid these problems, we introduce the propensity score.

The propensity score is the concept proposed by Rosenbaum and Rubin(1983) that enables covariate adjustment through a single variable into which the information of multiple covariate variables is aggregated; it is the coarsest one-dimensional balancing score\footnote{A balancing score $b(x)$ is a function of observed covariates $x$ such that the conditional distribution of $x$ given $b(x)$ is independent of assignments $z$; that is,
\begin{align}
x\indep z|b(x).
\end{align}
Balancing scores are not uniquely determined but various functions of $x$. The coarsest balancing score, i.e. the propensity score, is the function of any other balancing scores (Rosenbaum and Rubin, 1983).}.
In general propensity score analysis, a selection probability of a missing in the context of missing data analysis or a treatment assignment in the context of causal inference is usually used as a propensity score.
In this study, on the other hands, since what determines which of the potential outcome $Y_j$ is the group assignment, the selection probability of $D$ given the covariates $X$ and $W$ is regarded as the propensity score. Under the ignorability assumption (\ref{eq:ign}), we can estimate the conditional expectations as
\begin{align}
\begin{split}
&E_{X,W}\left[E_{D|X,W}\left[\frac{D}{E(D|X,W)}E(Y_0|X,D=1)\right]\right]\\
&\quad= E_{X,W}\left[E_{D|X,W}\left[\frac{D}{E(D|X,W)}\right]E(Y_0|X,D=1)\right]\\
&\quad= E[Y_0|X]\quad(X<c_1)\\
&E_{X,W}\left[E_{D|X,W}\left[\frac{D}{E(D|X,W)}E(Y_1|X,D=0)\right]\right]\\
&\quad= E_{X,W}\left[E_{D|X,W}\left[\frac{D}{E(D|X,W)}\right]E(Y_1|X,D=0)\right]\\
&\quad= E[Y_1|X]\quad(X>c_0)
\end{split}
.
\end{align}
The specific procedure to estimate as above is described in the next section.

\section{Estimation}
Estimation in the conventional RD designs has been considered as nonparametric estimation problem since the misspecification of the function form may cause bias in estimation of the causal effect (Hahn et al., 2001; Lee and Lemieux, 2010). Therefore we consider nonparametric estimation of $g_j(x)$, in particular, using the local linear regression model taking advantage of the fact that $X$ is one dimensional variable. 
Note that considering that the purpose of this research is estimation of counterfactual between the two thresholds and estimation of causal effect using it, it is sufficient to estimate even a regression function between thresholds. However, if the estimation target is limited to the interval between the thresholds, the bad boundary behavior of the kernel regression as above occur in the neighborhood of the thresholds. Since data exist outside the thresholds in this design, we use them to improve the stability of estimation; the target of estimation is not limited to the interval between the thresholds.

Now consider a nonparametric regression model $Y_i=g(X_i)+\varepsilon_i$, where $g(x)$ is a unknown smooth function. The local linear estimates of is $g(x)$ formed by minimizing
\begin{align}
\sum_{i=1}^n K_h(X_i-x)[Y_i-\bm{G}(X_i-x)^T\bm{\alpha}]^2 
\label{eq:llr}
\end{align}
where $K_h(X_i-x)=K(X_i-x/h)/h$ is the kernel weight with bandwidth $h$ and $\alpha \equiv (\alpha_0(x),\alpha_1(x))^T$, $G(X_i-x)\equiv(1,X_i-x)^T$. The estimated function is $\hat{g}(x)=\hat{\alpha}_0(x)$.
When complete data exists, $\bm{\alpha}$ solving the equation (\ref{eq:llr}) gives the correct regression function; however, actually, the presence of missing due to the design in our study make it biased in general. 

When focusing on estimate of $E(Y_0|X)$, data of $D=1$ is complete case for $X<c_1$. 
Consistent estimation of $E(Y_0|X)$ for $X<c_1$ can be implemented using data of $D=1$ and the inversed probability weighted (IPW) method or augmented inversed probability weighted (AIPW) method propsed by Wang et al.(2010), which is more robust than IPW. It is similar for estimate of $E(Y_1|X)$ for $X>c_0$ and data of $D=0$.
However, those estimation method ignore the other data ($D=0$ for $g_0(x)$ or $D=1$ for $g_1(x)$)  except in estimation of the selection probability model although those data are available.
Especially, the observed data of $D=0$ for $X<c_0$ ((c) in Figure \ref{fig:po}) and $D=1$ for $X>c_1$ ((d) in Figure \ref{fig:po}) including both the auxiliary variables and even the outcome can be used to estimate in the neighborhood of the thresholds $c_0$ and $c_1$, but the information of those is totally ignored. Those methods are not efficient in this respect.
Therefore we propose more efficient method which is capable of exploiting the information from even (c) or (d) in Figure \ref{fig:po}.

\subsection{Proposed doubly robust estimation}
We develop a new estimation method for the design of this study based on the AIPW kernel regression proposed by Wang et al. (2010).

As mentioned in the previous section, we consider covariate adjustment using propensity score under the ignorability assumption (\ref{eq:ign}) in order to implement unbiased estimation. Let $\pi_i=Pr(D_i=1|X_i,W_i)$ denote the data selection probability as propensity score and we assume a parametric model:
\begin{align}
\pi_i=\pi(X_i,W_i;\bm{\gamma}),
\label{eq:model_ps}
\end{align}
where $\bm{\gamma}$ is a finite dimensional parameter vector. 
This model can be specified as logit model or probit model, for example, and we estimate $\hat{\pi}_i=\pi(X_i,W_i;\hat{\bm{\gamma}})$ using $\hat{\bm{\gamma}}$, the maximum likelihood estimate of $\bm{\gamma}$. By weighting the units by the inverse of the estimated $\hat{\pi}_i$ or the inverse of the true selection probability $\pi_i$, if known, we obtain a inversed probability weighted (IPW) estimator.

Denote by $\delta_j(X_i,W_i)$ an arbitrary regression function of $X_i$ and $W_i$. To estimate $\delta_j(X_i,W_i)$ we postulate a parametric model 
\begin{align}
E(Y_{ji}|X_i,W_i)=\delta_j(X_i,W_i;\eta_j).
\label{eq:model_reg}
\end{align}
where $\eta_j$ is a finite dimensional parameter vector. We can estimate $\hat{\delta}_j(X_i,W_i;\hat{\eta}_j)$ by using $\hat{\eta}_j$, the estimate of $\eta_j$ obtained by the standard method such as OLS and by using data satisfying $Z=j$; $\hat{\eta}_0$ is estimated from the part as shown as (a) and (c) in Figure \ref{fig:po} and $\hat{\eta}_1$ is estimated from (b) and (d).

Now we define the estimating equation for $g_0(\cdot)$ as 
\begin{align}
\sum_{i \in N|X_i<c_1}[U_{IPW,i}^0(\bm{\alpha}^0)-A_i^0(\bm{\alpha}^0)] = 0,
\label{eq:pro0}
\end{align}
where
\begin{align}
\begin{split}
U^0_{IPW,i}(\bm{\alpha^0}) =& D_i\left[(1-Z_i) \frac{D_i}{\hat{\pi}_i}K_{h_0}(X_i-x)V_{0i}^{-1}\bm{G}(X_i-x)\left[Y_i-\bm{G}(X_i-x)\bm{\alpha^0}\right]\right] \\
& + (1-D_i)\left[(1-Z_i)\frac{1-D_i}{1-\hat{\pi}_i}K_{h_0}(X_i-x)V_{0i}^{-1}\bm{G}(X_i-x)\right.\\
& \qquad\qquad\qquad\qquad\qquad\qquad\qquad\qquad\qquad 
\left. \times\left[Y_i-\bm{G}(X_i-x)\bm{\alpha^0}\right]\right] \\
\label{eq:U_pro}
\end{split}
\end{align}
\begin{align}
\begin{split}
A^0_i(\bm{\alpha^0}) =& D_i\left[\left((1-Z_i)\frac{D_i}{\hat{\pi}_i}-1\right)K_{h_0}(X_i-x)V_{0i}^{-1}\bm{G}(X_i-x)\right.\\
&\qquad\qquad\qquad\qquad\qquad\qquad\qquad\qquad\qquad  \left. \times\left[\hat{\delta}_0(X_i,W_i;\hat{\eta}_0)-\bm{G}(X_i-x)\bm{\alpha^0}\right]\right]\\
&+ (1-D_i)\left[\left((1-Z_i)\frac{1-D_i}{1-\hat{\pi}_i}-1\right)K_{h_0}(X_i-x)V_{0i}^{-1}\bm{G}(X_i-x)\right.\\
&\qquad\qquad\qquad\qquad\qquad\qquad\qquad\qquad\qquad  \left. \times
\left[\hat{\delta}_0(X_i,W_i;\hat{\eta}_0)-\bm{G}(X_i-x)\bm{\alpha^0}\right]\right]
\label{eq:A_pro}
\end{split}
\end{align}
and for $g_1(\cdot)$ as
\begin{align}
\sum_{i \in N|X_i>c_0}\left[U_{IPW,i}^1(\bm{\alpha}^1)-A_i^1(\bm{\alpha}^1)\right] =0
\label{eq:pro1}
\end{align}
where
\begin{align}
\begin{split}
U^1_{IPW,i}(\bm{\alpha^1}) &=D_i \left[ Z_i\frac{D_i}{\hat{\pi}_i}K_{h_1}(X_i-x)V_{1i}^{-1}\bm{G}(X_i-x)\left[Y_i-\bm{G}(X_i-x)\bm{\alpha^1}\right]\right] \\
&\qquad + (1-D_i)\left[ Z_i\frac{1-D_i}{1-\hat{\pi}_i}K_{h_1}(X_i-x)V_{1i}^{-1}\bm{G}(X_i-x)\right.\\
&\qquad\qquad\qquad\qquad\qquad\qquad\qquad\qquad\qquad\qquad\quad
\left. \times
\left[Y_i-\bm{G}(X_i-x)\bm{\alpha^1}\right]\right] \\
A^1_i(\bm{\alpha^1}) & =D_i\left[Z_i \left(\frac{D_i}{\hat{\pi}_i}-1\right)K_{h_1}(X_i-x)V_{1i}^{-1}\bm{G}(X_i-x)\right.\\
&\qquad\qquad\qquad\qquad\qquad\qquad\qquad\qquad\quad
\left. \times\left[\hat{\delta}_1(X_i,W_i;\hat{\eta}_1)-\bm{G}(X_i-x)\bm{\alpha^1}\right]\right]\\
&\qquad + (1-D_i)\left[\left(Z_i\frac{1-D_i}{1-\hat{\pi}_i}-1\right)K_{h_1}(X_i-x)V_{1i}^{-1}\bm{G}(X_i-x)\right.\\
&\qquad\qquad\qquad\qquad\qquad\qquad\qquad\qquad\quad
\left. \times\left[\hat{\delta}_1(X_i,W_i;\hat{\eta}_1)-\bm{G}(X_i-x)\bm{\alpha^1}\right]\right]\\
\end{split}
\end{align}
with $\alpha^j=(\alpha_0^j(x),\alpha_1^j (x))$ solving equation (\ref{eq:pro0}) or equation (\ref{eq:pro1}) is the local linear estimator of $g_j(x)$, $V_{ji}=V[G(X_i-x)^T \alpha^j;\zeta_j]$ with a known working variance function $V(\cdot, \cdot)$ and an unknown finite dimensional parameter $\zeta_j$. 
The consistency of the estimation is guaranteed even if $V_j$ is arbitrarily decided under certain conditions (Hoshino, 2009). If we estimate $\zeta_0$ based on the data, we can use the inverse probability weighted moment equations $\sum_{l=1}^n D_l\hat{\pi}_l^ {-1} V_{0l}^{(1)} \left[\left\{Y_l - \hat{\alpha}^0_{0,l} ( \zeta_0 ) \right\}^2 - V \left\{ \hat{\alpha}^0_{0,l} ( \zeta_0 ) , \zeta_0 \right\} \right] = 0$, where $V _ l ^ { ( 1 ) } = \partial V \left\{ \hat { \alpha}^0_{0,l} (\zeta_0 ) ; \zeta_0 \right\} / \partial \zeta_0 $, and $\hat { \alpha } _ { l } ( \zeta_0 ) = \left\{ \hat { \alpha } _ { 0 , l } ( \zeta_0 ) , \hat { \alpha } _ { 1 , l } ( \zeta_0 ) \right\} ^ { T }$ solve (\ref{eq:pro0}) with $x = X _ { l } , l = 1 , \dots , n$. We can estimate $\zeta_1$ in a similar way. The estimated conditional expectation function is $\hat{g}_j(x)=\hat{\alpha}_0^j(x)$. The first term of equation (\ref{eq:pro0}) and (\ref{eq:pro1}) is what constitutes the IPW estimation equation as $\sum U_{IPW,i}^j(\alpha^j)=0$ and the second term $A_i^j(\alpha^j)$ is called an augmented term.

\begin{figure}[htbp] 
 \begin{center}
  \begin{minipage}[c]{1\linewidth}
   \centering
   \includegraphics[width=0.8\linewidth]{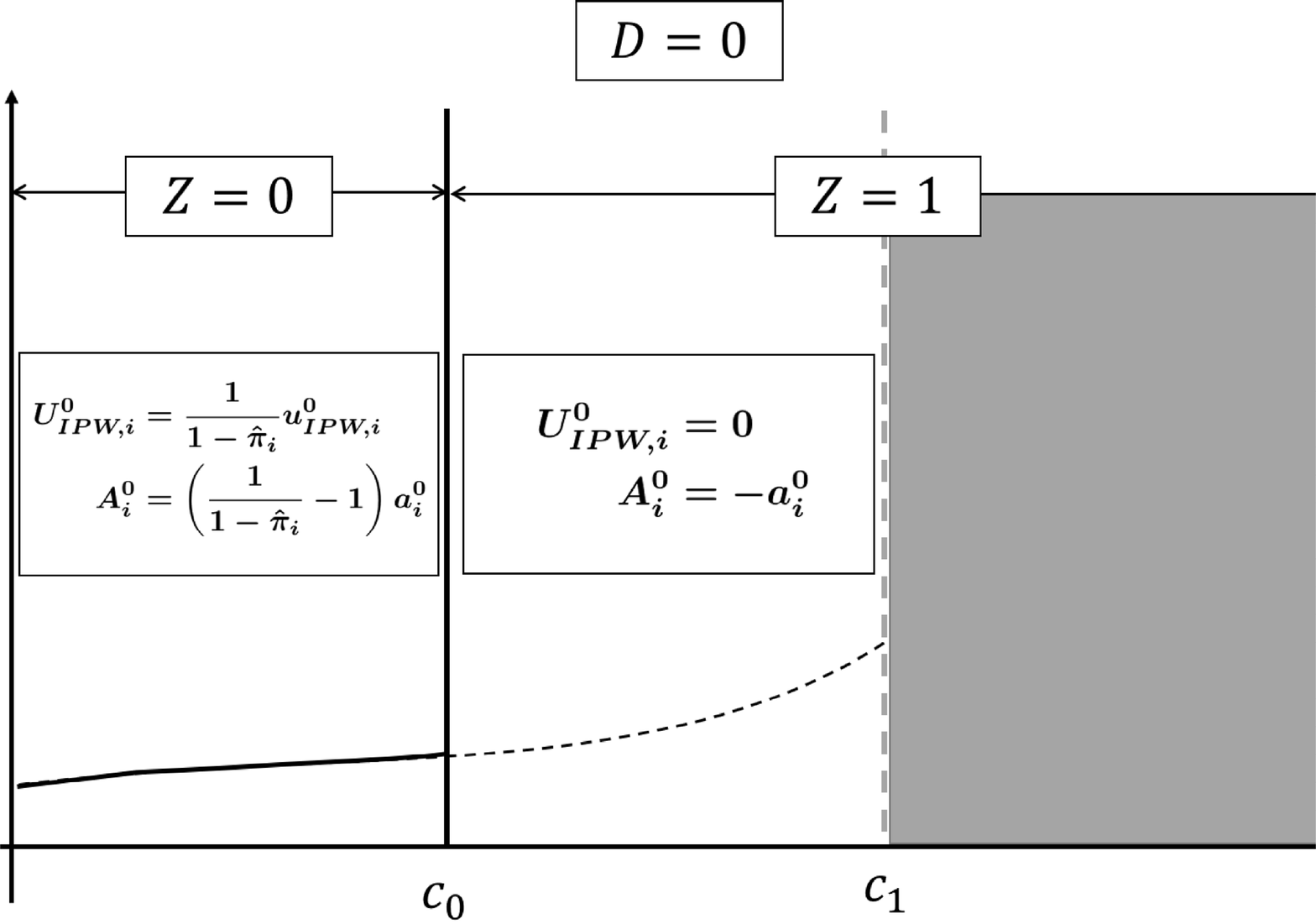}
   \subcaption{}
 \end{minipage}\\
       \begin{minipage}{1\linewidth}
        \vspace{7mm}
      \end{minipage} \\
 \begin{minipage}[c]{1\linewidth}
 \centering
 \includegraphics[width=0.8\linewidth]{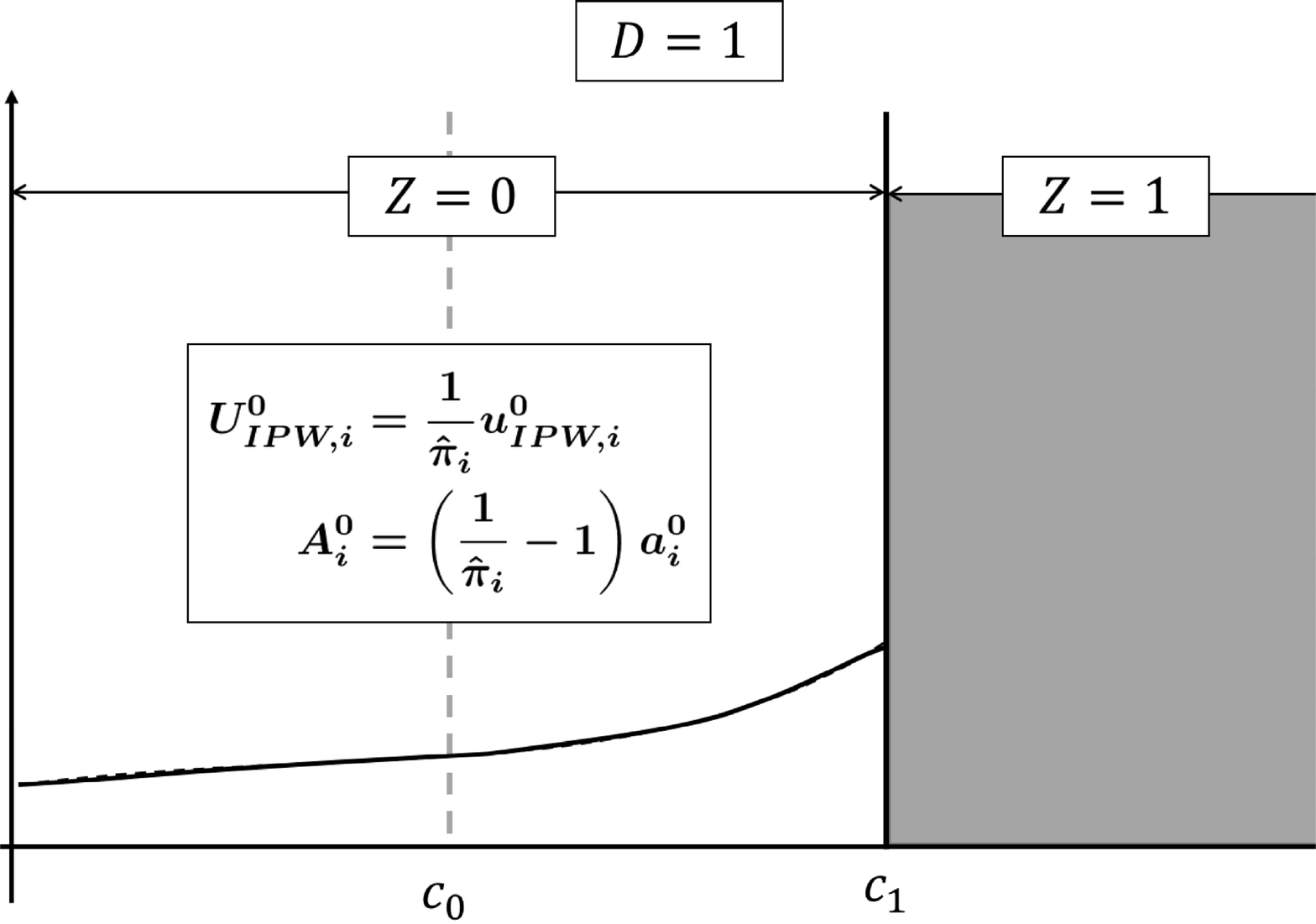}
    \subcaption{}
  \end{minipage} 
 \end{center}
　\caption{Depending on $D_i$ and $Z_i$ how subjects are included in the proposed estimating equation for $g_0(x)$. 
 $u^0_{IPW,i}=K_{h_0}(X_i-x)V_{0i}^{-1}\bm{G}(X_i-x)\left[Y_i-\bm{G}(X_i-x)\bm{\alpha^0}\right]$
 and
 $a^0_i=K_{h_0}(X_i-x)V_{0i}^{-1}\bm{G}(X_i-x)\left[\delta_0(X_i,W_i)-\bm{G}(X_i-x)\bm{\alpha^0}\right]$. The top panel for $D_i=0$ and the bottom panel is for $D_i=1$. 
}
 %\raggedright
 \label{fig:model}
\end{figure}

We inevestigate properties of the estimation equations focusing on for $g_0(\cdot)$. These equations allow us to use data of $D=0$ in addition to $D=1$. When $D_i=1$, the first terms in the right hand side in equation (\ref{eq:U_pro}) and (\ref{eq:A_pro}) are left and the scond terms are equal to 0, and thus this estimating equation is equal to the one proposed by Wang et al. (2010). When $D_i=0$, the second terms are left and the first terms are equal to 0.  For unit $i\in N_0$, $Z_i$ differs depending on either $X_i\leq c_0$ or $X_i>c_0$. If $X_i\leq c_0$, i.e. $Z_i=0$, since complete data including outcomes exists, outcomes and covariates can be included in the estimation as well as $D_i = 1$ in the form changing weight to $1-\hat{\pi}_i$. On the other hand, if $X_i>c_0$, i.e. $Z_i=1$, potential outcome $Y_{0i}$ is regarded as missing but covariates are obtained. In this case, whereas $U_{IPW,i}$ is equal to 0 by $1-Z_i=0$, the augmented term $A_{i}$ is left with weight $-1$. Therefore the information of covariates can be exploited. 
Now if only data of $D=0$ is used to estimate the parameter $\eta_0$ in $\delta_0(X_i,W_i; \eta_0)$, since the potential outcomes $Y_0$ are obtained only for $X_i\leq c_0$, applying estimated parameters to units satisfying $X_i>c_0$ is an extrapolation and it is not desirable. However in equation (\ref{eq:pro0}) units satisfying $D_i=1$ and $X_i>c_0$ are weighted by the selection probability and included in addition to the data of $D_i=0$ and thus it can be interpreted as an interpolation.
Figure \ref{fig:model} shows in what forms units are included in the estimation depending on $D_i$ and $Z_i$.

The estimators solving equation (\ref{eq:pro0}) or (\ref{eq:pro1}) also have the double-robustness similar to other AIPW estimators including the one proposed by Wang et al.(2010). The estimator is consistent when either of the two following conditions is satisfied (but not necessarily both): (i) the selection probability model is correctly specified, and (ii) the regression function of all covariates is correctly specified. The double-robustness is prooved in the next section.

\subsection{Bandwidth selection}
Appropriate choice of bandwidth is an important issue in kernel regression. The least squares cross validation (LSCV) is one of the most widely used bandwidth selection methods (Li and Racine, 2007). Let $\hat{g}_{0,-i}(X_i)$ and $\hat{g}_{1,-i}(X_i)$ denote the leave-one-out local linear estimator of $g_{0}(X_i)$ and $g_{1}(X_i)$. $\hat{g}_{0,-i}(X_i)$ is the solution in the equation 
\begin{align}
\sum_{l\not=i, l \in N|X_l<c_1}[U_{IPW,l}^0(\bm{\alpha}^0)-A_l^0(\bm{\alpha}^0)] =0
\end{align}
where
\begin{align}
\begin{split}
U^0_{IPW,l}(\bm{\alpha^0}) &= D_l\left[(1-Z_l) \frac{D_l}{\hat{\pi}_l}K_{h_0}(X_l-X_i)V_{0l}^{-1}\bm{G}(X_l-X_i)\left[Y_l-\bm{G}(X_l-X_i)\bm{\alpha^0}\right]\right] \\
&\qquad + (1-D_l)\left[(1-Z_l)\frac{1-D_l}{1-\hat{\pi}_i}K_{h_0}(X_l-x)V_{0l}^{-1}\bm{G}(X_l-X_i)\right.\\
&\qquad\qquad\qquad\qquad\qquad\qquad\qquad\qquad\qquad\qquad\quad
\left. \times\left[Y_l-\bm{G}(X_l-X_i)\bm{\alpha^0}\right]\right] \\
A^0_l(\bm{\alpha^0}) & = D_l\left[\left((1-Z_l)\frac{D_l}{\hat{\pi}_l}-1\right)K_{h_0}(X_l-X_i)V_{0l}^{-1}\bm{G}(X_l-X_i)\right.\\
&\qquad\qquad\qquad\qquad\qquad\qquad\qquad\qquad\quad
\left. \times\left[\delta_0(X_l,W_l)-\bm{G}(X_l-X_i)\bm{\alpha^0}\right]\right]\\
& \qquad + (1-D_l)\left[\left((1-Z_l)\frac{1-D_l}{1-\hat{\pi}_l}-1\right)K_{h_0}(X_l-X_i)V_{0l}^{-1}\bm{G}(X_l-X_i) \right.\\
&\qquad\qquad\qquad\qquad\qquad\qquad\qquad\qquad\quad
\left. \times \left[\delta_0(X_l,W_l)-\bm{G}(X_l-X_i)\bm{\alpha^0}\right]\right]
\end{split}
\end{align}
and $\hat{g}_{1,-i}(X_i)$ solves a equation similar to the above.
The LSCV method choose the bandwith minimizing a function of bandwidth $h$,
denoted as $LSCV_j(h)$, as the optimal bandwidth. $LSCV_0(h)$ and $LSCV_1(h)$ are respectively defined as 
\begin{align}
LSCV_0(h) = \frac{1}{\sum_{i \in N|X_i<c_1}(1-Z_i)}\sum_{i \in N|X_i<c_1}(1-Z_i)(Y_i-\hat{g}_{0,-i}(X_i))^2
\end{align}
and 
\begin{align}
LSCV_1(h) = \frac{1}{\sum_{i \in N|X_i>c_0}Z_i}\sum_{i \in N|X_i<c_1}Z_i(Y_i-\hat{g}_{1,-i}(X_i))^2.
\end{align}

Therefore the optimal bandwidth is defined as  
\begin{align}
h_{j,opt}\equiv\argmin_h LSCV_j(h).
\end{align}

See Li and Racine (2007) for the mathematical details of the local linear cross validation.

\section{Asymptotic Properties\label{sec:AP}}
In this section, we describe the asymptotic properties of the estimator proposed in this paper. We can investigate it in a similar way to Wang et al.(2010).
Throughout this section we assume the following: (I) $n\rightarrow \infty$, $h\rightarrow 0$, and $nh\rightarrow \infty$: (II) $x$ is in the interior of the support of $X$: (III) the regularity conditions: (i) $g(\cdot)$ and the densitiy function of X, $f_X(\cdot)$ satisfy the smoothness assumptoions of Fan et al. (1996); (ii) the right hand side of the estimating equation are twice continuously differentiable with respect to $\alpha$ at a target point $x$ and second derivatives are uniformly bounded. 

The proposed doubly robust (DR) local linear estimator of $g_j(x)$ is $\hat{g}_{j,DR}(x)$ solving equation (\ref{eq:pro0}) or (\ref{eq:pro1}) and this asymptotic limit is denote by $\tilde{g}_{j,DR}(x)$. The proposed DR kernel estimating equations (\ref{eq:pro0}) or (\ref{eq:pro1}) should have a sequence of solutions $(\hat{\alpha}^j_{0,DR}(x),\hat{\alpha}^j_{1,DR}(x))$ at $x$ such that as the sample size $n\rightarrow \infty$, and the sequence converges in probability to a vector $(\tilde{\alpha}^j_{0,DR}(x), \tilde{\alpha}^j_{1,DR}(x))$, of which the first component $\tilde{\alpha}^j_{0,DR}(x)$ is denoted by $\tilde{g}_{j,DR}(x)$, and $\tilde{g}_{0,DR}(x)$ satisfies
\begin{align}
\begin{split}
&E\left[(1-Z)\frac{D}{\tilde{\pi}}V_0^{-1}\{\tilde{g}_{0,DR}(x); \tilde{\zeta}_0\}\left[Y_0-\tilde{g}_{0,DR}(x)\right]|X=x\right]\\
&\quad+E\left[D\left((1-Z)\frac{D}{\tilde{\pi}}-1\right)V_0^{-1}\{\tilde{g}_{0,DR}(x); \tilde{\zeta}_0\}\left[\tilde{\delta}_0(X,W)-\tilde{g}_{0,DR}(x)\right]|X=x\right]\\
&\quad+E\left[(1-Z)\frac{1-D}{1-\tilde{\pi}}V_0^{-1}\{\tilde{g}_{0,DR}(x); \tilde{\zeta}_0\}\left[Y_0-\tilde{g}_{0,DR}(x)\right]|X=x\right]\\
&\quad+E\left[(1-D)\left((1-Z)\frac{1-D}{1-\tilde{\pi}}-1\right)V_0^{-1}\{\tilde{g}_{0,DR}(x); \tilde{\zeta}_0\}\right.\\
&\qquad\qquad\qquad\qquad\qquad\qquad\qquad\qquad\qquad
\left. \times\left[\tilde{\delta}_0(X,W)-\tilde{g}_{0,DR}(x)\right]|X=x\right]=0,
\label{eq:con_DR0}
\end{split}
\end{align}
and $\tilde{g}_{1,DR}(x)$ satisfies
\begin{align}
\begin{split}
&E\left[Z\frac{D}{\tilde{\pi}}V_1^{-1}\{\tilde{g}_{1,DR}(x); \tilde{\zeta}_1\}\left[Y_1-\tilde{g}_{1,DR}(x)\right]|X=x\right]\\
&\quad+E\left[D\left(Z\frac{D}{\tilde{\pi}}-1\right)V_1^{-1}\{\tilde{g}_{1,DR}(x); \tilde{\zeta}_1\}\left[\tilde{\delta}_1(X,W)-\tilde{g}_{1,DR}(x)\right]|X=x\right]\\
&\quad+E\left[Z\frac{1-D}{1-\tilde{\pi}}V_1^{-1}\{\tilde{g}_{1,DR}(x); \tilde{\zeta}_1\}\left[Y_1-\tilde{g}_{1,DR}(x)\right]|X=x\right]\\
&\quad+E\left[(1-D)\left(Z\frac{1-D}{1-\tilde{\pi}}-1\right)V_1^{-1}\{\tilde{g}_{1,DR}(x); \tilde{\zeta}_1\}\right.\\
&\qquad\qquad\qquad\qquad\qquad\qquad\qquad\qquad\qquad
\left. \times\left[\tilde{\delta}_1(X,W)-\tilde{g}_{1,DR}(x)\right]|X=x\right]=0,
\label{eq:con_DR1}
\end{split}
\end{align}
where $\tilde{\pi}=\pi(X_i,W_i; \tilde{\gamma})$ and $\tilde{\gamma}$ is the probability limit of $\hat{\gamma}$, and $\tilde{\delta}_j(X,W)=\delta_j(X,W; \tilde{\eta_j})$ and $\tilde{\eta_j}$ is the probability limit of $\hat{\eta_j}$.
Theorem \ref{thm:con} provides the consistency of the proposed estimator under certain conditions.

\begin{thm}
Under the ignorability assumption, the probability limit $\tilde{g}_{j,DR}(x)$ defined in equation (\ref{eq:con_DR0}) and (\ref{eq:con_DR1}) satisfies $\tilde{g}_{j,DR}(x)=g_{j,DR}(x)$, that is, consistent estimator of $g_{j,DR}(x)$ when either of the following conditions is satisfied;
\begin{enumerate}
\renewcommand{\labelenumi}{(\alph{enumi})}
 \setlength{\leftskip}{0.3cm}
 \item The selection probability $\hat{\pi}_i$ in the DR estimating equation (\ref{eq:pro0}) is replaced by the true selection probability $\pi_i$ or by the estimated $\hat{\pi}_i=\pi_i(X_i,W_i; \hat{\gamma})$ with $\hat{\gamma}$ which is computed under the correctly specified model.
 \item The regression function $\delta_j(X,W)$ satisfies $\delta_j(X,W)=E(Y_j|X,W)$ or $\delta_j(X,W)=\delta_j(X,W;\hat{\eta_j})$ with $\hat{\eta_j}$ which is computed under the correctly specified model.
\end{enumerate}
 \label{thm:con}
\end{thm}

Theorem \ref{thm:con} shows the double-robustness of the proposed estimator as mentioned previously. The proof of Theorem \ref{thm:con} about $\hat{g}_{0,DR}$ is shown in what follows.

\begin{proof}
Under the the strong ignorability condition (\ref{eq:strig}) and the ignorability assumption (\ref{eq:ign}) and , equation (\ref{eq:con_DR1}) can be rewritten as
\begin{align}
\begin{split}
&E\left[\left[Y_0-\tilde{g}_{0,DR}(x)\right]|X=x\right]
+E\left[\left((1-Z)\frac{D}{\tilde{\pi}}-1\right)\left[Y_0-\tilde{\delta}_0(X,W)\right]|X=x\right]\\
&\quad+E\left[\left[Y_0-\tilde{g}_{0,DR}(x)\right]|X=x\right]
+E\left[\left((1-Z)\frac{1-D}{1-\tilde{\pi}}-1\right)\right.\\
&\qquad\qquad\qquad\qquad\qquad\qquad\qquad\qquad\qquad\qquad\qquad
\left. \times\left[Y_0-\tilde{\delta}_0(X,W)\right]|X=x\right]=0
\label{eq:con_DR2}
\end{split}
\end{align}
When the true seletion probability is known or the seletion probability model (\ref{eq:model_ps}) is correctly specfied, that is,  $\tilde{\pi}=E(D|X,W)$, or the regression function (\ref{eq:model_reg}) is correctly specified, that is, $\tilde{\delta}_0(X,W)=E(Y_0|X,W)$, the second and fourth terms of eqaution (\ref{eq:con_DR2}) are 0. Hence eqaution (\ref{eq:con_DR1}) is equal to 
\begin{align}
E[[Y_0-\tilde{g}_{0,DR}(x)]|X=x]=0
\end{align}
Therefore we have $\tilde{g}_{0,DR}(x)=g_0(x)$, that is, $\hat{g}_{0,DR}$ is a consistent estimator of $g_0(x)$.
\end{proof}
Theorem \ref{thm:con} about $\hat{g}_{1,DR}$ can be easily proved in a similar way.

Next we invesitgate the asymptotic distribution of the proposed estimator.
Theorem \ref{thm:ad} shows the asymptotic bias and variance of the proposed estimator.
\begin{thm}
Assume that 
\begin{enumerate}
\renewcommand{\labelenumi}{(\roman{enumi})}
 \setlength{\leftskip}{0.3cm}
 \item the selection probability $\hat{\pi}_i$ in the estimating equation (\ref{eq:pro0}) is computed under a model (\ref{eq:model_ps}) or replaced by fixed probabilities $\hat{\pi}^*_i=\hat{\pi}^*(X_i,W_i)$;
 \item the regression function $\delta_j(X,W)$ in the estimating equation (\ref{eq:pro0}) and (\ref{eq:pro1}) is a known function or replaced by the function $\delta_j(X,W;\hat{\eta_j})$ with $\hat{\eta_j}$ which is estimated on units with observed outcomes;
  \item $Pr(D=1|X,W)>a>0$ for some constant $a$ with probability $1$ in a neighborhood $X=x$ ;
 \item The ignorability assumption (\ref{eq:ign}) and assumption(I)-(III) hold
\end{enumerate}
In addition to the above assumptions, consider the two conditions;
\begin{enumerate}
\renewcommand{\labelenumi}{(\Alph{enumi})}
 \setlength{\leftskip}{0.3cm}
 \item The selection probability $\hat{\pi}_i$ in the DR estimating equation (\ref{eq:pro0}) is replaced by the true selection probability $\pi_i$ or by the estimated $\hat{\pi}_i=\pi_i(X_i,W_i; \hat{\gamma})$ with $\hat{\gamma}$ which is computed under the correctly specified model.
 \item The regression function $\delta_j(X,W)$ is a known function or replaced by the function $\delta_j(X,W;\hat{\eta_j})$ with $\hat{\eta_j}$ which is estimated on units with observed outcomes under the correctly specified model.
\end{enumerate}
If either (A) or (B) holds at least, but necessarily not both, then 
\begin{align}
\sqrt{nh}\left\{\hat{g}_{0,DR}-g_0(x)-\frac{1}{2}h^2\{g_0(x)\}^{\prime\prime}c_2(K)+o(h^2)\right\}\longrightarrow N\left(0,W^0_{DR}(x)\right)
\label{eq:ad}
\end{align}
where 
\begin{align}
\begin{split}
W^0_{DR}(x) = b_K(x) &E\left[ \left[ D \left\{ \frac{(1-Z)D}{\tilde{\pi}(X,W)}\left(Y_0-g_0(X)\right)-\left(\frac{(1-Z)D}{\tilde{\pi}(X,W)}-1\right)\right.\right.\right. \\ 
& \left.\left.\left.\qquad\qquad\qquad\qquad\qquad\qquad\qquad\qquad\quad
\times \left(\tilde{\delta}_0(X,W)-g_0(X)\right)\right\} \right.\right. \\ 
&\quad+ \left.\left.(1-D) \left\{ \frac{(1-Z)(1-D)}{1-\tilde{\pi}(X,W)}\left(Y_0-g_0(X)\right)\right.\right.\right. \\ 
& \left.\left.\left.\qquad
-\left(\frac{(1-Z)(1-D)}{1-\tilde{\pi}(X,W)}-1\right)\left(\tilde{\delta}_0(X,W)-g_0(X)\right) \right\} \right]^2|X=x\right]
\end{split}
\label{eq:av}
\end{align}
and
\begin{align}
\sqrt{nh}\left\{\hat{g}_{1,DR}-g_1(x)-\frac{1}{2}h^2\{g_1(x)\}^{\prime\prime}c_2(K)+o(h^2)\right\}\longrightarrow N\left(0,W^1_{DR}(x)\right)
\label{eq:ad1}
\end{align}
where 
\begin{align}
\begin{split}
W^1_{DR}(x) = b_K(x) &E\left[ \left[ D \left\{ \frac{ZD}{\tilde{\pi}(X,W)}\left(Y_1-g_1(X)\right)-\left(\frac{ZD}{\tilde{\pi}(X,W)}-1\right)\right.\right.\right. \\ 
& \left.\left.\left.\qquad\qquad\qquad\qquad\qquad\qquad\qquad\qquad\quad
\times \left(\tilde{\delta}_1(X,W)-g_1(X)\right)\right\} \right.\right. \\ 
&\quad+ \left.\left.(1-D) \left\{ \frac{Z(1-D)}{1-\tilde{\pi}(X,W)}\left(Y_1-g_1(X)\right)\right.\right.\right. \\ 
& \left.\left.\left.\qquad
-\left(\frac{Z(1-D)}{1-\tilde{\pi}(X,W)}-1\right)\left(\tilde{\delta}_1(X,W)-g_1(X)\right) \right\} \right]^2|X=x\right]
\end{split}
\label{eq:av1}
\end{align}
with $f_X(x)$ is the density function of $X$, $b_K(x)\equiv \int K^2(s)ds/f_X(x)$ and $c_2(K)\equiv \int s^2K(s)ds$ .
\label{thm:ad}
\end{thm}

Theorem \ref{thm:ad} shows that the asymptotic bias of the proposed estimator is of order $O(h^2)$, and the variance of it is of order $O(1/nh)$, and additionally, it is independent of the working variance $V(\cdot)$ in the proposed DR kernel estimating equations (\ref{eq:pro0}). 
A proof of Theorem \ref{thm:ad} is provided in the Appendix. 

\section{Estimation of an optimal threshold}
The principal aim of this study are expanding the conventional RD design the purpose of which is evaluating the causal effect at the discontinuous point to estimate counterfactual between the two thresholds and to enable evaluation of the causal effect at arbitrary points between the thresholds themselves. In this section, moreover, we propose to estimate optimal thresholds in terms of cost effectiveness as an application of this study. Our position here is to support policy makers' decisions.

In general, it is considered desirable to target as many subjects as possible if special interventions yield better results. However, in practice, special interventions require more costs than regular interventions and the intervention practitioners (e.g. governments or companies) need to bear additional costs. For above reasons, they limit subjects by setting uniform criteria and that is why the RD design is useful in many cases. Considering such background, it is obvious that the question of where to set the threshold to maximize cost performance is one of the most important issues for practitioners. In the web marketing example described above, who pay the expense for the privilege of greater membership or for the coupons are the companies providing such services and it is easy to imagine that they cannot help limiting the target customers due to budgetary reasons. Setting criteria to maximize the return on investment in this example is an important management challenge.

In what follows, we describe how to optimize the threshold using the estimated counterfactuals. Attention should be given to the fact that the following discussion is based on the presupposition that outcomes and cost are measured by the same unit, which is supposed to be money in most cases. Indeed, there are cases where outcomes and costs are variables of different measurement especially in political cases, however this problem has been dealt with in another research area, namely cost benefit analysis. We regard this problem as a issue deviating from the range of this research and do not deal with it here.

We postulate that the optimal thresholds can be estimated by maximization of the function of a threshold $c$ representing the total benefit obtained in the treatment group and the control group minus the additional costs with constraint subject to $c_0<c<c_1$; that is,  
\begin{align}
\begin{split}
&\max_{c\in[c_0,c_1]}(E[Y_0|X<c]Pr(X<c)+E[Y_1|c<X]Pr(X>c)-m(c))\\
&\quad=\max_{c\in[c_0,c_1]}\left(\int^c_{-\infty} g_0(x)f_X(x)dx+\int_c^\infty g_1(x)f_X(x)dx-m(c) \right)
\end{split}
,
\end{align}
where $m(c)$ is a known function of a threshold $c$ that represents the additional cost of treatment and $f_X(x)$ is a probability density function of $X$. The benefits obtained from $X<c_0$ and $X>c_1$ are constant for every $c\in [c_0, c_1]$, thus, practically, we need to consider only maximization of the total benefits and costs across the thresholds. Therefore, the optimal threshold can be defined as
\begin{align}
\begin{split}
c_{opt}&\equiv \argmax_{c\in[c_0,c_1]}(E[Y_0|c_0<X<c]Pr(c_0<X<c)+E[Y_1|c<X<c_1]Pr(c<X<c_1)-m(c))\\
& = \argmax_{c\in[c_0,c_1]}\left(\int^c_{c_0} g_0(x)f_X(x)dx+\int_c^{c_1} g_1(x)f_X(x)dx-m(c) \right)
\end{split}
.
\label{eq:copt_def}
\end{align}
Since it is assumed that the same intervention is performed for all subjects, it is considered reasonable to assume that the additional cost per unit is constant. Therefore, the cost function can be defined as 
\begin{align}
m(c)\equiv \int_c^{c_1}MC(x) f_X(x)dx
\end{align}
where $MC(c)$ is the additional cost per unit when threshold is set to $c$. Using this definition and equation (\ref{eq:ce}), the objective function of optimization is
\begin{align}
\begin{split}
E[Y_0&|X<c]Pr(X<c)+E[Y_1|c<X]Pr(X>c)-m(c)\\
&=\int_{c_0}^c g_0(x)f_X(x)dx + \int_c^{c_1}g_1(x)f_X(x)dx - \int_c^{c_1}MC(x)f_X(x)dx\\
&=\int_{c_0}^{c_1} g_0(x)f_X(x)dx - \int_c^{c_1}g_0(x)f_X(x)dx + \int_c^{c_1}\{g_1(x)-MC(x)\}f_X(x)dx\\
&=\int_{c_0}^{c_1} g_0(x)f_X(x)dx - \int_c^{c_1}\{\tau(x)-MC(x)\}f_X(x)dx
\end{split}
\end{align}
When the intervention providers are beneficiaries at the same time such as the web marketing example mentioned above and $\forall x\in [c_0,c_1]$, $\tau(x)-MC(x)<0$, in other words, $\max\tau(x)<MC(x)$ is satisfied, the objective function is monotonically decreasing and hence the optimal threshold is estimated as $c_{opt}=c_0$. However, this result means that the additional benefit due to the treatment (i.e. the causal effect) is less than the additional cost at any point the intervention does not pay off  and implies that the validity of the intervention itself might have to be reviewed from the viewpoint of cost effectiveness.

The practical estimator of the optimal threshold is $\hat{c}_{opt}$ solving equation (\ref{eq:copt_def}) with $g_j(x)$  replaced by $\hat{g}_j(x)$ estimated in the method proposed in Section 3 and either the true probability density function $f_X(x)$, if known as prior information, or an estimator of it $\hat{f_X}(x)$ estimated by the kernel density estimation, for instance. 

\section{Simulations}
In this section, we describe simulation conducted to investigate the properties of the proposed estimator in the finite samples. We evaluate our proposed estimator by comparing it with IPW local linear estimator and the naive local linear estimator.
The IPW local linear estimator solves the first terms of equation (\ref{eq:pro0}) and (\ref{eq:pro1}) $\sum U^j_{IPW,i}=0$ using the data of either $D=0$ or $D=1$. The naive local linear estimator solves equation formed by specification of $\pi$ in the IPW estimating equation to be 1.
We generate 100 data sets and estimate using each data set under the following conditions to evaluate the estimators from some viewpoints. 
First, in order to study the efficiency of the proposed estimator, we performed three types of estimation for each data set using the proposed estimator with all units, the IPW local linear estimator with either of the groups except in the estimation of selection probabilities and the naive local linear estimator with complete case.  Note that model specifications here in the IPW and proposed estimator are correct.
Next, for the evaluation of the robustness of the estimators, compare the results in the following four cases; (i) the selection probability model in the IPW estimation is incorrect; (ii) the selection probability model in the proposed estimation is incorrect; (iii) the regression model in the proposed estimation is incorrect; (iv) both of the models of $\pi$ and $\delta$ are incorrect. 
Finally, we examine dependency on the settings of the distribution of the running variable by generating the running variable from either the normal distribution or the log normal distribution.
We evaluate the estimation results by comparing mean integrated squared error (MISE) limited to between the two thresholds defined as $\int_{c_0}^{c_1} \{ \hat{g}(x)-g(x)\}f_X(x)dx$. 
We use the LSCV method to choose the optimal bandwidth as described in Section 4.4.

\begin{figure}[htbp]
 \begin{minipage}{0.5\hsize}
  \begin{center}
\includegraphics[width=\linewidth]{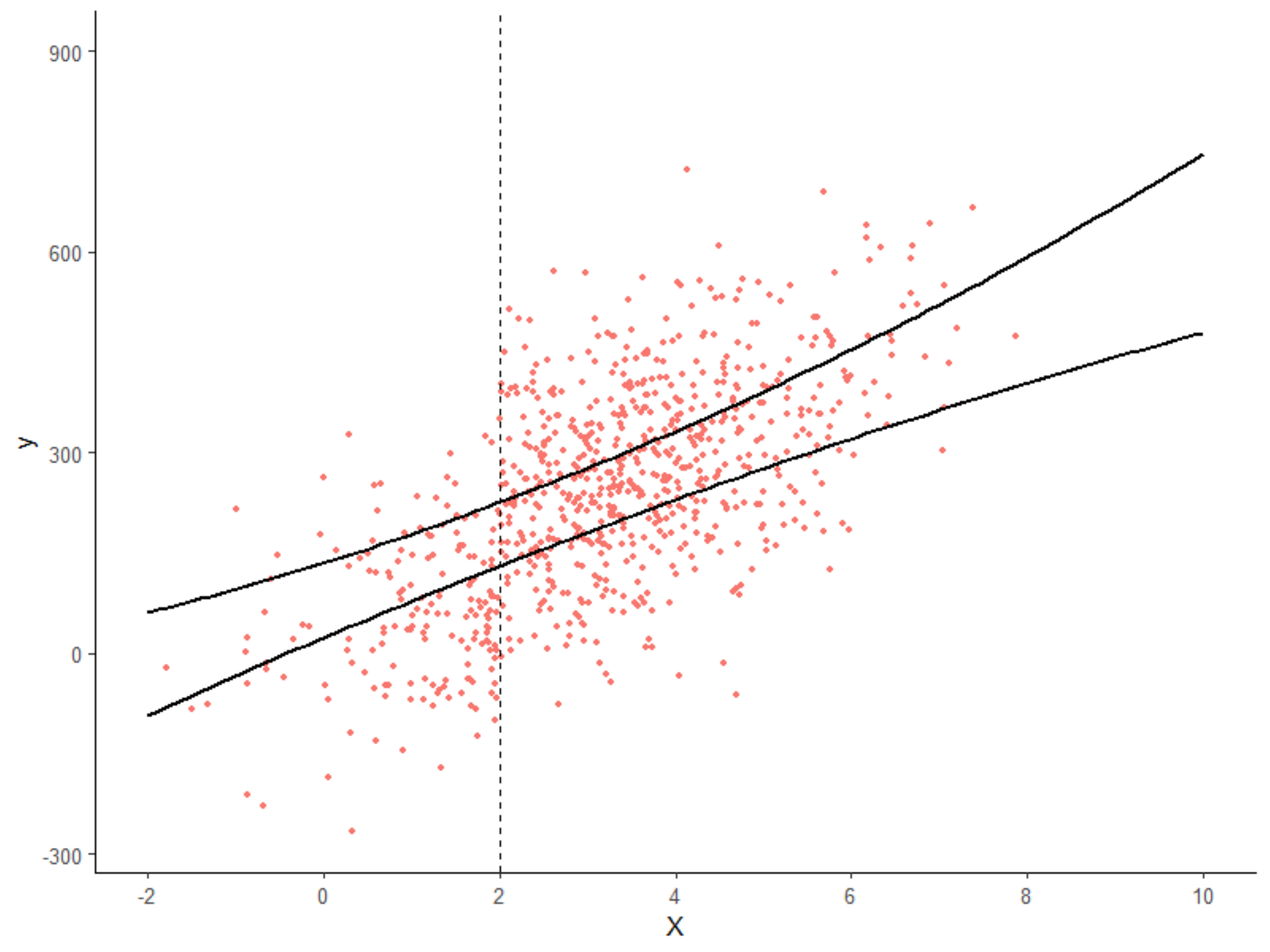}
  \end{center}
  \subcaption{}
  \label{fig:one}
 \end{minipage}
 \begin{minipage}{0.5\hsize}
  \begin{center}
\includegraphics[width=\linewidth]{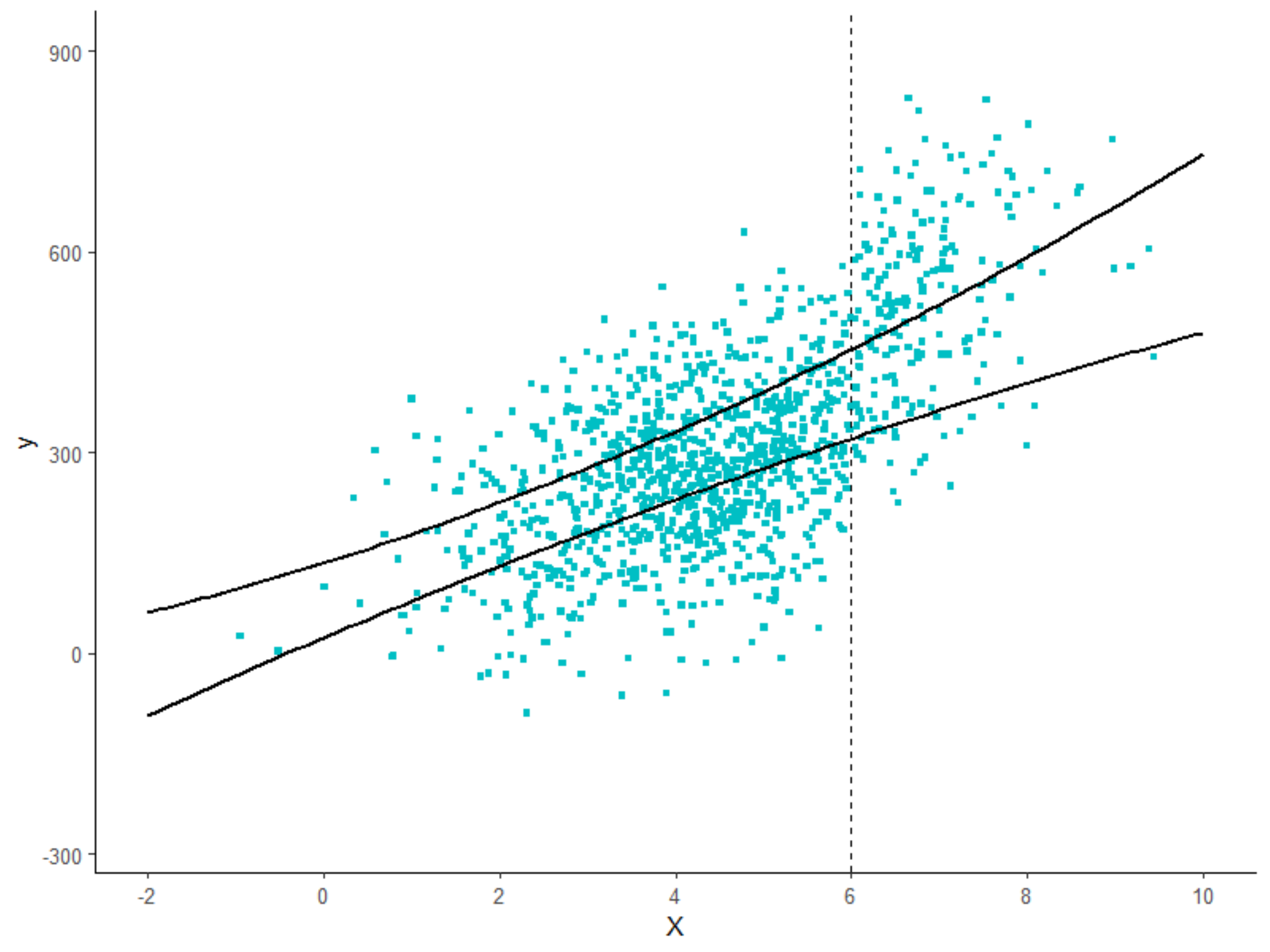}
  \end{center}
  \subcaption{}
  \label{fig:two}
 \end{minipage}\\
 \begin{center}
 \begin{minipage}{0.8\hsize}
  \begin{center}
\includegraphics[width=0.9\linewidth]{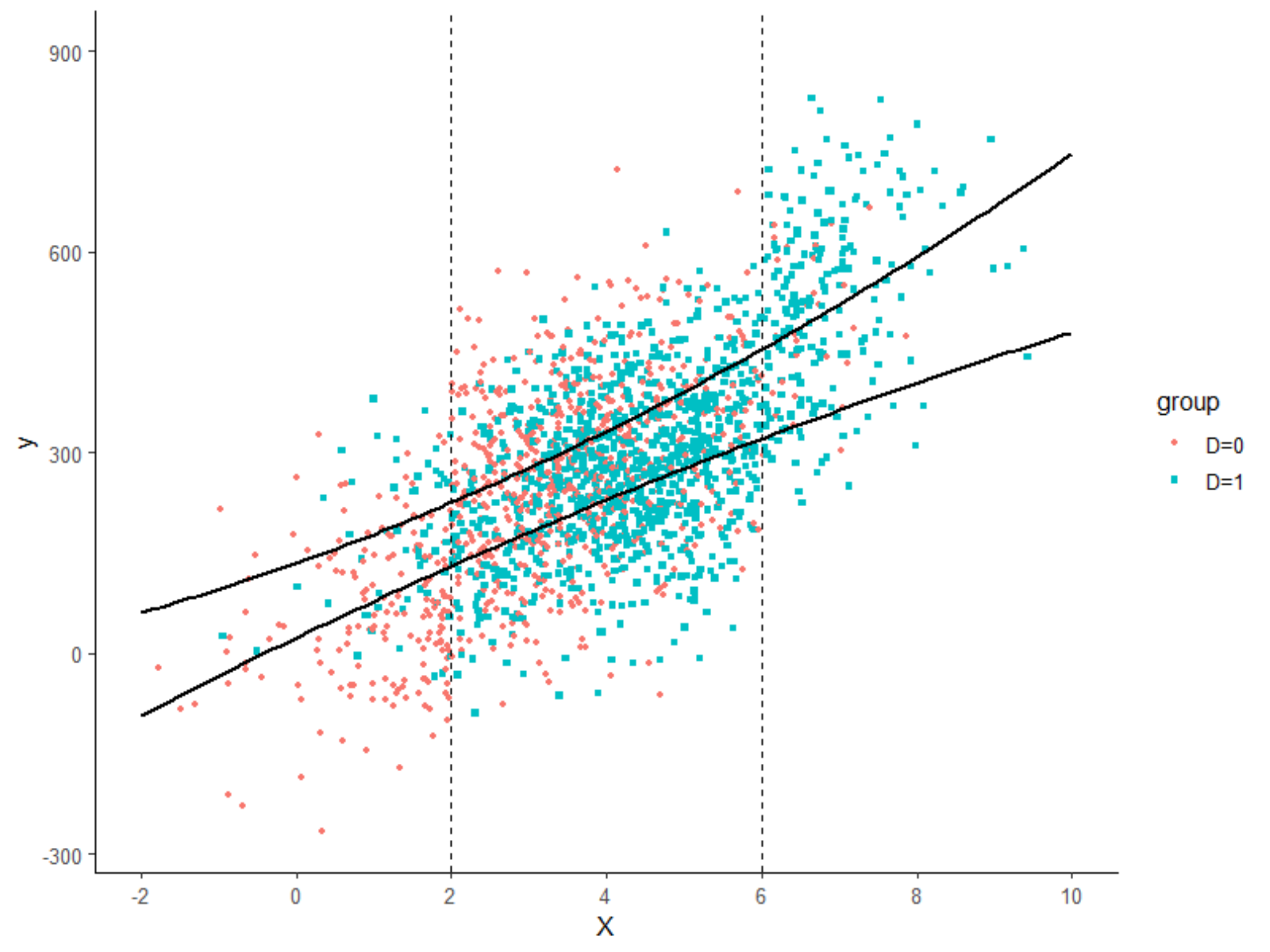}
  \end{center}
  \subcaption{}
  \label{fig:one}
 \end{minipage}
 \end{center}
 \caption{Scatter plots of one of the generated data sets. The left panel of the top is for $D=0$ and the right panel is for $D=1$ and the bottom panel is a scatter (X, Y) plot on the same plane. In each panel the upper black line indicates $E(Y_1|X)=E_{W|X}(E(Y_1|X,W))$ and the lower one indicates $E(Y_0|X)=E_{W|X}(E(Y_0|X,W))$}
 \label{fig:plot}
\end{figure}

In what follows, describe the data generating process.
The running variable $X$ is generated from a normal distribution with mean 4 and variance $\sigma^2=1.7^2$.
We assume that in this simulation the covariates $\boldsymbol{W}$ other than $X$ is 2-dimensional and correlated with $X$ to induce selection bias. Thus we generate $\boldsymbol{W}=(w_1,w_2)^T$ according to a model: $\boldsymbol{W}=\boldsymbol{\eta_0}+\boldsymbol{\eta_1} X+\bm{\xi}$, where $\boldsymbol{\eta_0}$ and $\boldsymbol{\eta_1}$ are $2 \times 1$ parameter vectors and $\boldsymbol{\eta_0}=(-1.5, 2.4)^T$ and $\boldsymbol{\eta_1}=(0.6, 0.4)^T$, $\bm{\xi}=(\xi_1,\xi_2 )^T$ is the disturbance term generated from normal distribution with mean $0$ and $\sigma^2=4$ independently. For the data assignment probability $\pi_i$ we postulate the logit model
\begin{align}
logit(\pi_i )=\gamma_0+\gamma_1 X_i+ \gamma_2 w_{1i}+\gamma_3 w_{2i},
\label{eq:pi}
\end{align}
where $\gamma_0=0.8$, $\gamma_1=0.5$, $\gamma_2=2$ and $\gamma_3=-0.8$. Then the data assignment indicator $D_i$ is sampled from Bernoulli distribution with probability $\pi_i$. Following $D_i$, the treatment assignment $Z_i$ is determined by the function $Z_i=1\left(X_i>c_{d_i}\right)$, with the lower thresholds $c_0=2$ and the upper thresholds $c_1=6$. Finally we generate the observed outcomes $Y_i=Z_iY_{1i}+(1-Z_i)Y_{0i}$, where 
\begin{align}
Y_{ji}=\beta^j_0 + \beta^j_1 X_i+ \beta^j_2 X_i^2 + \beta^j_3 w_{1i} + \beta^j_4 w_{2i}+\varepsilon_{ji},~~~       \varepsilon_{ji} \overset{i.i.d.}{\sim}  N(0,10^2)
\label{eq:delta}
\end{align}
with $(\beta^0_0, \beta^0_1, \beta^0_2, \beta^0_3, \beta^0_4)=(0, 16, -1, 42, 36)$ and $(\beta^1_0, \beta^1_1, \beta^1_2, \beta^1_3, \beta^1_4)=(80, -2, 2, 40, 48)$. Figure \ref{fig:plot} shows a scatter plot of the $(X,Y)$ from one of the generated data sets with the lines that indicate $E(Y_j|X)=E_{W|X}(E(Y_j|X,W))$. 

\begin{figure}[htbp]
  \begin{center}
\includegraphics[width=8cm]{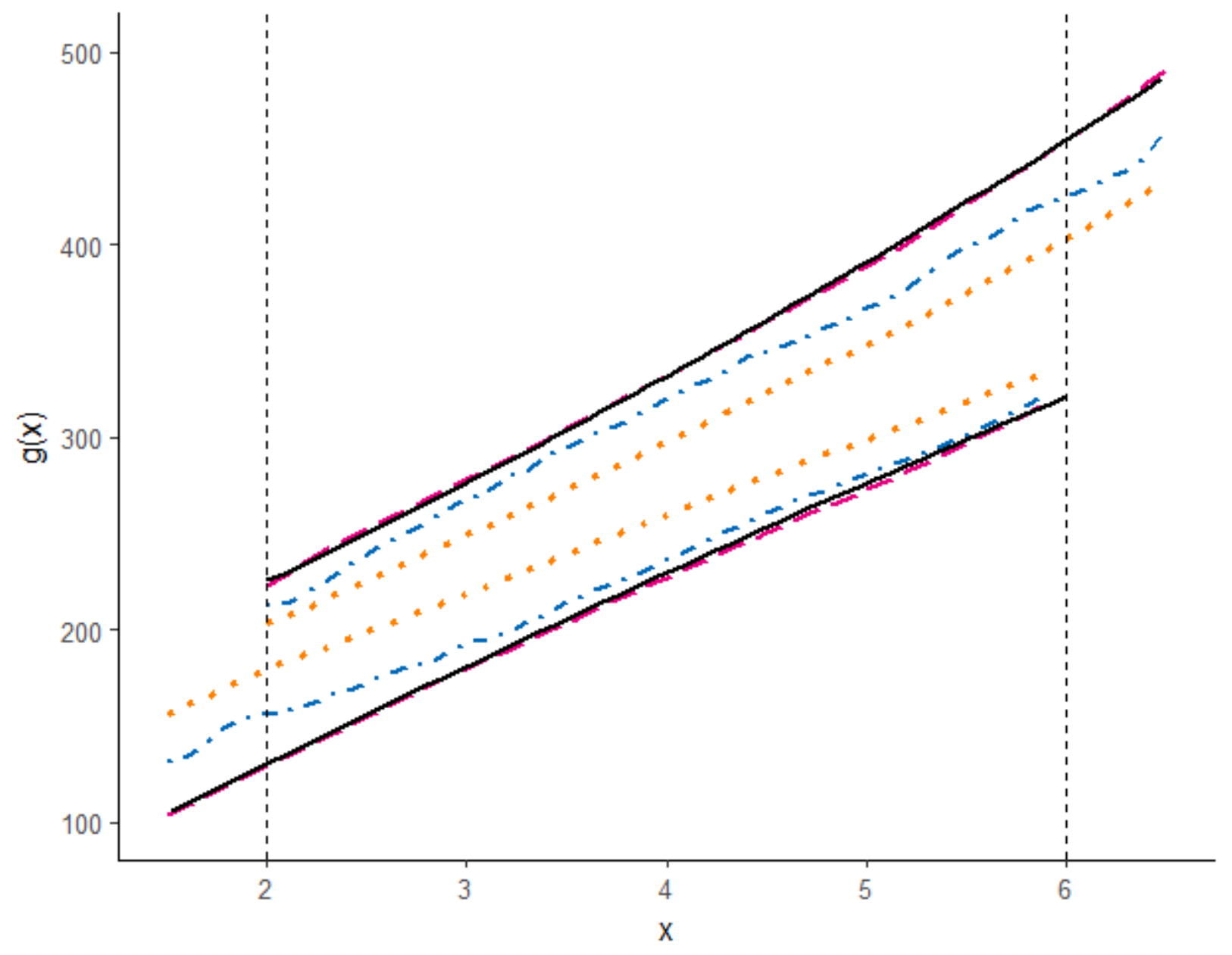}
  \end{center}
 \caption{Estimated nonparametric functions of $g_0(x)$ and  $g_0(x)$ using the naive, IPW and DR estimation methods. The black solid lines are true $g_0(x)$ and  $g_1(x)$, the red dashed lines are estimated functions using the DR estimation, the blue dashed-dotted lines are estimated functions using the IPW estimation and the orange dotted lines are estimated functions using the naive local linear estimation.
 }
 \label{fig:line}
\end{figure}

In what follows we report the results when sample size $n=2000$ . Figure \ref{fig:line} and Table \ref{tab:res1} show the result of the naive, IPW and DR local linear estimators of $g_0(x)$ and $g_1(x)$. Table \ref{tab:res1} summarizes the MISEs of the naive, IPW and DR local linear estimators of $g_0(x)$ and $g_1(x)$ as the performance with correct models. The naive local linear estimates have much larger MISEs than the IPW and AIPW local linear estimates for both of $g_0(x)$ and $g_1(x)$. The DR local linear estimates have smaller MISEs than the IPW local linear estimates. For instance, the DR local linear estimate has approximately 59\% gain in MISE efficiency in comparison with the IPW local linear estimate in estimation of $E(Y_0|X)$.

\begin{table}[htbp]
  \begin{center}
    \begin{tabular}{c}

      % 1
      \begin{minipage}[t]{0.45\hsize}
        \begin{center}
    \caption{MISEs of the naive, IPW and DR local linear estimates of $g_0(x)$ and $g_1(x)$ with correct models}
  \begin{tabular}{lcccc} \hline\hline
        \multicolumn{1}{l}{}& \multicolumn{2}{c}{MISE} \\

      \multicolumn{1}{l}{}& \multicolumn{1}{c}{$g_0(x)$} & \multicolumn{1}{c}{$g_1(x)$}\\\hline
 %      & MISE  & MISE  \\\hline \hline
    Naive & 829.2  & 1060.5  \\
  %  \multicolumn{5}{c}{correctly specified}\\
	IPW & 314.8  & 740.2  \\
    DR & 130.5 & 683.3 \\ \hline
  \end{tabular}
  \label{tab:res1}
        \end{center}
      \end{minipage}
      
\begin{minipage}{0.02\hsize}
\hspace{0.5mm}
      \end{minipage}

      % 2
      \begin{minipage}[t]{0.5\hsize}
        \begin{center}
   \caption{MISEs of the IPW and DR local linear estimates of $g_0(x)$ using $\hat{\pi}$ and/or $\hat{\delta}$ computed under incorrectly specified models}
  \begin{tabular}{lcc} \hline\hline
 %     \multicolumn{1}{l}{}& \multicolumn{2}{c}{$E(Y_0|X)$} \\
       & MISE  \\\hline 
 %   \multicolumn{3}{c}{incorrectly specified}\\
    IPW($\pi$ wrong) & 818.3  \\
    DR($\pi$ wrong) & 181.4  \\
    DR($E(Y_0|X,W)$ wrong) & 546.2  \\
    DR(both wrong) &  607.6 \\ \hline
  \end{tabular}
  \label{tab:res2}
        \end{center}
      \end{minipage}

    \end{tabular}
  \end{center}
\end{table}

Since it is expected that results for $g_0(x)$ and $g_1(x)$ have similar tendencies from the theory and the results shown in Table \ref{tab:res1}, we focus on estimation of $g_0(x)$ in the following simulations. Next consider the case that $\pi$ and/or $\delta$ of the IPW and DR are incorrectly specified as described above. The incorrect model of $\pi$ is specified as the model (\ref{eq:pi}) without the $w_{1}$ term and the incorrect model of $\delta$ is specified as the model (\ref{eq:delta}) without the $X$ squared term. Table \ref{tab:res2} shows the results with incorrectly specified models. The DR estimate with a misspecified $\pi$ has relatively close to the DR estimate with correct models and it is better than the IPW estimate with correct $\pi$. The DR estimate with a misspecified $\delta$ is not as good as the DR estimate with a misspecified $\pi$, however its MISE is still better than the Naive estimate and the IPW estimate with an incorrect $\pi$, and naturally better than the DR estimates when both the model of $\pi$ and $\delta$ are misspecified. 

\section{Discussion}
In this paper we proposed a new framework of the regression discontinuity designs for estimation of two conditional expectation functions of potential outcomes, i.e. counterfactuals, between two thresholds by using multiple groups which have difference thresholds. We considered how to realize estimation of them in the two cases with and without selection bias. We showed that we can simply estimate them in the absence of selection bias but cannot generally in the presence of it using the normal estimation method such as the naive local linear regression. In order to estimate consistently and to make the best use of the available data, we proposed the new estimator based on the AIPW kernel estimator with the ignorability assumption. We showed that the proposed estimator has double-robustness and it can exploit the auxiliary information of covariates from even subjects with missing outcomes. In finite samples, the proposed estimator is more efficient compared with the naive local linear estimator and the IPW kernel estimator and has  the double-robustness property.

One of the concerns about this study is whether a regression model with a mixture of two data sets as a population is meaningful even if the ignorability (\ref{eq:ign}) is assumed. If we wish to infer the results for a more general population, that is possible when covariates including a running variable are obtained from the more general population and we can assume that which group subjects belong to is determined by the covariates. 

In addition, in this paper we have chosen nonparametric regression to estimate conditional expectation functions for some reasons, however parametric regressions are also used in many empirical RD designs. If parametric conditional expectation functions are postulated, counterfactual can be estimated at any point on a running variable by extrapolation with the estimated parameters and covariates from beyond the observation range without using the proposed estimation method.

There is room for the further development of this research. First of all, we should apply the proposed method to real data to confirm its usefulness in empirical cases. As for the theoretical side, in this paper we proposed our method focusing on limited case in some respects.  We have considered only the case with two groups, however our method can be extended to cases with three or more groups. Another important topic of future study is an extension to the fuzzy RD design not limited to the sharp RD design.

\newpage

\appendix

\renewcommand{\theequation}{A.\arabic{equation} }
\setcounter{equation}{0}

\section{Proof of Theorem 2: the asymptotic bias and variance of the proposed estimator}
The proof of Theorem 2 follows similar arguments as those in Appendix of Wang et al.(2010). We focus our proof on the estimator of $g_0(x)$. For any interior point $x$, reparameterize $\alpha$ as $\{g_0(x),hg_0'(x)\}^T$ and denoted by $g_{0,true}(x)$ the true value of $g_0(x)$, $\alpha_0=\{g_{0,true}(x),hg_{0,true}'(x)\}^T$ and $\hat{\alpha}_{DR}(x)$ the solution of the local linear DR kernel estimating equations (\ref{eq:pro0}). The asymptotic results hold when the parameters $(\gamma, \eta)$ in $\pi$ and $\delta$ are estimated at the $\sqrt{n}$-rate, or the probability limit of $(\hat{\gamma}, \hat{\eta})$ is used in the DR kernel estimating equations (\ref{eq:pro0}). Denote by $(\tilde{\gamma}, \tilde{\eta})$ the probability limit of $(\hat{\gamma}, \hat{\eta})$, and let $\tilde{\pi}(X_i,W_i) = \pi(X_i,W_i; \tilde{\gamma})$, $\tilde{\delta}_0(X_i, W_ i) = \delta_0(X_i,W_i; \tilde{\eta})$. We focus our proof on the case assuming that $(\tilde{\gamma}, \tilde{\eta})$ are known. In addition, we assume that the variance parameter $\zeta$ in the working variance $V$ is known.

A Taylor expansion of the local linear DR kernel estimating equations (\ref{eq:pro0}) gives
\begin{align}
\sqrt{nh} \left\{\hat{\alpha}_{DR}(x)-\alpha_0\right\} = -\sqrt{nh}\left\{\Gamma_{n,\delta} \left(\alpha_{*}\right)\right\}^{-1} \Lambda_{n,\delta } \left(\alpha_0\right)
\end{align}
where $\alpha_\ast$ is between $\hat{\alpha}_{DR}(x)$ and $\alpha_{0}$,
\begin{align}
\begin{split}
\Lambda_{n,\delta}(\alpha) =& \frac{1}{n} \sum_{i \in N|X_i<c_1} \left[D_i\left\{\frac{(1-Z_i)D_i}{\widetilde{\pi} \left(X_i, W_i\right)} K_h\left(X_i-x\right) V_i^{-1}(x, \alpha)G\left(X_i-x \right)\right.\right.\\
&\qquad  \left.\left.
\times \left[Y_i- G \left(X_i - x \right)^T \alpha\right]
- \left\{ \frac{(1-Z_i)D_i}{\widetilde{\pi} \left(X_i, W_i\right)} - 1 \right\} K_h\left(X_i-x\right) V_i^{-1}(x, \alpha)\right.\right.\\
&\qquad\qquad\qquad\qquad\qquad\qquad  \left.\left.
\times  G\left(X_i-x \right) \left[\widetilde{\delta}_0(X_i,W_i)- G \left(X_i - x \right)^T \alpha\right]\right\}\right.\\
&\quad + (1-D_i)\left.\left\{ \frac{(1-Z_i)(1-D_i)}{1-\widetilde{\pi} \left(X_i, W_i\right)} K_h\left(X_i-x\right) V_i^{-1}(x, \alpha)G\left(X_i-x \right)\right.\right.\\
&\qquad  \left.\left.
\times \left[Y_i- G \left(X_i - x \right)^T \alpha\right]
-\left\{ \frac{(1-Z_i)(1-D_i)}{1-\widetilde{\pi} \left(X_i, W_i\right)} - 1 \right\} K_h\left(X_i-x\right)\right.\right.\\
&\qquad\qquad\qquad\qquad  \left.\left.
\times V_i^{-1}(x, \alpha) G\left(X_i-x \right) \left[\widetilde{\delta}_0(X_i,W_i)- G \left(X_i - x \right)^T \alpha\right]\right\}\right]
\end{split}
\end{align}
where $n=\#\{ i | i \in N, X_i<c_1 \}$, $V_i^{-1}(x, \alpha)=V\{G(X_i-x)^T\alpha;\zeta_0\}$ and $\Gamma_{n,\delta}(\boldsymbol{\alpha}) = \partial \boldsymbol {\Lambda}_{n, \delta } ( \boldsymbol { \alpha } )/\partial \boldsymbol { \alpha }^T$.

We consider the following two situations:
\begin{enumerate}
\renewcommand{\labelenumi}{\arabic{enumi}).}
 \item When model (\ref{eq:model_ps}) for the selection probability $\pi_{i0}$ is correctly specified, i.e. $\widetilde{\pi}\left( X_i, \boldsymbol{W}_i\right) = \pi_{i0} \left(X_{i}, \boldsymbol{W}_i \right)$; 
 \item When model (\ref{eq:model_reg}) for $E(Y_0|X,W)$ is correctly specified, i.e. $\widetilde{\delta}_0 \left(X_i, \boldsymbol{W}_i \right) = E(Y_{0i}| X_i, \boldsymbol{W}_i )$
\end{enumerate}
As shown in Section 5, $\hat{\alpha}_{DR}(x)$ converges to $\alpha_{0}$ when either of the above conditions holds. Therefore, $\alpha_\ast \overset{P}{\longrightarrow} \alpha_0$. We first show that under either of the above situations, we have
\begin{align}
\boldsymbol {\Gamma}_{n,\delta} \left(\boldsymbol{\alpha}_{*}\right) \stackrel{P}{\longrightarrow}-f_{X}(x) V^{-1}\{g_0(x)\}\boldsymbol{D}(K)
\label{eq:proconv}
\end{align}
where $D(K)$ is a $2 \times 2$ matrix with the ($j,k$)th element $c_{j+k-2}(K) \times h^{(j+k-2)}$ and $c_r(K) = \int s^r K(s)ds$

First consider situation 1)., i.e., when $\widetilde{\pi}\left( X_i, \boldsymbol{W}_i\right) = \pi_{i0} \left(X_{i}, \boldsymbol{W}_i \right)$. The second and fourth terms of $\Lambda_{n,\delta}(\alpha)$, i.e. the augmentation terms, has mean $0$ under MAR (\ref{eq:strig}) and the ignorability assumption (\ref{eq:ign}). It follows that $\Lambda_{n,\delta} \left(\boldsymbol{\alpha}_*\right) = \boldsymbol{\Lambda}_n\left( \boldsymbol{\alpha}_0 \right) + \boldsymbol{o}_p(1)$, where $\boldsymbol{\Lambda}_n$ is formed by 
\begin{align}
\begin{split}
\Lambda_n(\alpha) =& \frac{1}{n} \sum_{i \in N|X_i<c_1} \left\{D_i\frac{(1-Z_i)D_i}{\widetilde{\pi} \left(X_i, W_i\right)} K_h\left(X_i-x\right) V_i^{-1}(x, \alpha)G\left(X_i-x \right)\right.\\
&\qquad\qquad  \left.
\times \left[Y_i- G \left(X_i - x \right)^T \alpha\right]
+ (1-D_i)\frac{(1-Z_i)(1-D_i)}{1-\widetilde{\pi} \left(X_i, W_i\right)} K_h\left(X_i-x\right) \right.\\
&\qquad\qquad\qquad\qquad\qquad\qquad  \left.
\times V_i^{-1}(x, \alpha)G\left(X_i-x \right) \left[Y_i- G \left(X_i - x \right)^T \alpha\right]\right\}%\\
\end{split}
\end{align}
Hence $\Gamma_{n,\delta}\left(\boldsymbol{\alpha}_*\right) = \Gamma_n \left( \boldsymbol {\alpha}_0 \right) + \boldsymbol {o}_p( 1 )$, where $\Gamma_n(\boldsymbol{\alpha}) = \partial \boldsymbol {\Lambda}_n( \boldsymbol { \alpha } )/\partial \boldsymbol { \alpha }^T$. Therefore $\Gamma_{n,\delta}\left(\boldsymbol{\alpha}_*\right)$ has the same probability limit as $\Gamma_n\left(\boldsymbol{\alpha}_*\right)$. Under MAR (\ref{eq:strig}) and the ignorability assumption (\ref{eq:ign}), simple calculation shows that 
\begin{align}
\begin{split}
\Gamma_n\left(\alpha_*\right) & = - E \left[ K_h(X-x) V^{-1}\left(x,\alpha_0\right) G (X-x)G(X-x)^T \right] + o_p(1) \\ 
& =-f_X(x)V^{-1}\{g_0(x)\}D(K)+o_p( 1 ),
\end{split}
\end{align}
and thus (\ref{eq:proconv}) holds for $\Gamma_{n,\delta}\left(\boldsymbol{\alpha}_*\right)$.

Next consider situation 2)., i.e., when $\widetilde{\delta}_0 \left(X_i, \boldsymbol{W}_i \right) = E(Y_{0i}| X_i, \boldsymbol{W}_i )$. Rewrite $\Lambda_{n,\delta}(\alpha)$ as
\begin{align}
\begin{split}
\Lambda_{n,\delta}(\alpha) =& \frac{1}{n} \sum_{i \in N|X_i<c_1} \left[ D_i \left\{ \frac{(1-Z_i)D_i}{\widetilde{\pi} \left(X_i, W_i\right)} K_h\left(X_i-x\right) V_i^{-1}(x, \alpha) G\left(X_i-x \right)\right.\right.\\
&\left.\left. \qquad\qquad\qquad\qquad\qquad\qquad\qquad\qquad\qquad\qquad\qquad
\times \left[Y_i- \widetilde{\delta}_0(X_i,W_i) \right]\right.\right.\\
&\left.\left. + K_h\left(X_i-x\right) V_i^{-1}(x, \alpha)G\left(X_i-x \right) \left[\widetilde{\delta}_0(X_i,W_i)- G \left(X_i - x \right)^T \alpha\right] \right\}\right.\\
&\left. + (1-D_i)\left\{ \frac{(1-Z_i)(1-D_i)}{1-\widetilde{\pi} \left(X_i, W_i\right)} K_h\left(X_i-x\right) V_i^{-1}(x, \alpha)G\left(X_i-x \right)\right.\right.\\
&\left.\left. \qquad\qquad\qquad\qquad\qquad\qquad\qquad\qquad\qquad\qquad\qquad
\times  \left[Y_i- \widetilde{\delta}_0(X_i,W_i) \right]\right.\right.\\
&\left.\left. + K_h\left(X_i-x\right) V_i^{-1}(x, \alpha)G\left(X_i-x \right) \left[\widetilde{\delta}_0(X_i,W_i)- G \left(X_i - x \right)^T \alpha\right] \right\} \right]
\end{split}
\end{align}
One can easily see the first and third terms on the right hand side has mean 0. It follows that
\begin{align}
\begin{split}
\Lambda_{n,\delta}(\alpha) &= \frac{1}{n} \sum_{i \in N|X_i<c_1} K_h\left(X_i-x\right) V_i^{-1}(x, \alpha)G\left(X_i-x \right)\\
&\qquad\qquad\qquad\qquad\qquad\qquad
\times \left[\widetilde{\delta}_0(X_i,W_i)- G \left(X_i - x \right)^T \alpha\right]+\boldsymbol{o}_p(1).
\end{split}
\end{align}
Differentiating it with respect to $\boldsymbol{\alpha}$ shows that $\Gamma_{n,\delta}\left(\boldsymbol{\alpha}_*\right) = \Gamma_n \left( \boldsymbol {\alpha}_0 \right) + \boldsymbol {o}_p( 1 )$. Therefore, (\ref{eq:proconv})
still holds in this situation.

Therefore, when either the $\pi$ or $\delta$ model is correctly specified, we have
\begin{align}
\sqrt {nh} \left\{\hat{\alpha}_{DR} (x) - \alpha_0 \right\} = \left\{f_X(x) V^{-1} \{g_0(x)\} D(K) \right\}^{-1}\sqrt{nh} \Lambda_{n,\delta} \left(\alpha_0\right) + o_p(1)
\label{eq:discon}
\end{align}
Write $\boldsymbol { \Lambda } _ { n , \delta } \left( \boldsymbol { \alpha } _ { 0 } \right) = \boldsymbol { \Lambda } _ { 1 n , \delta } \left( \boldsymbol { \alpha } _ { 0 } \right) - \boldsymbol { \Lambda } _ { 2 n , \delta } \left( \boldsymbol { \alpha } _ { 0 } \right) + \boldsymbol { \Lambda } _ { 3 n , \delta } \left( \boldsymbol { \alpha } _ { 0 } \right)$, where

\begin{align}
\begin{split}
\Lambda _ { 1 n ,\delta}\left(\alpha_0\right) = \frac{1}{n} \sum_{i \in N|X_i<c_1} \left[ \frac{(1-Z_i)D_i}{\widetilde{\pi} \left(X_i, W_i\right)} K_h\left(X_i-x\right) V_i^{-1}(x, \alpha)\left[Y_i- g_0(X_i)\right]G\left(X_i-x \right)\right.\\
\left.+ \frac{(1-Z_i)(1-D_i)}{1-\widetilde{\pi} \left(X_i, W_i\right)} K_h\left(X_i-x\right) V_i^{-1}(x, \alpha)\left[Y_i- g_0(X_i)\right]G\left(X_i-x \right)\right],
\end{split}
\end{align}

\begin{align}
\begin{split}
\Lambda _ { 2 n , \delta } \left( \alpha _ { 0 } \right) &= \frac{1}{n} \sum_{i \in N|X_i<c_1} \left[D_i\left\{ \frac{(1-Z_i)D_i}{\widetilde{\pi} \left(X_i, W_i\right)} - 1 \right\} K_h\left(X_i-x\right) V_i^{-1}(x, \alpha) \right.\\
&\left. \qquad\qquad\qquad\qquad\qquad\qquad\qquad\qquad
\times\left[\widetilde{\delta}_0(X_i,W_i)- g_0(X_i)\right]G\left(X_i-x \right)\right.\\
&\left.\qquad\qquad + (1-D_i) \left\{ \frac{(1-Z_i)(1-D_i)}{1-\widetilde{\pi} \left(X_i, W_i\right)} - 1 \right\} K_h\left(X_i-x\right) V_i^{-1}(x, \alpha) \right.\\
&\left. \qquad\qquad\qquad\qquad\qquad\qquad\qquad\qquad
\times \left[\widetilde{\delta}_0(X_i,W_i)- g_0(X_i)\right]G\left(X_i-x \right)\right],
\end{split}
\end{align}
and
\begin{align}
\Lambda_{3n,\delta}\left(\alpha_0 \right) = \frac{1}{n} \sum_{i \in N|X_i<c_1} K_h\left(X_i-x\right) V_i^{-1}(x, \alpha) \left[g_0(X_i) - G \left(X_i - x \right)^T \alpha \right] G\left(X_i-x \right)
\end{align}

One can easily see that $\Lambda_{1n,\delta}\left(\alpha_0 \right)$ and $\Lambda_{2n,\delta}\left(\alpha_0 \right)$ have mean $0$ when either $\pi_i$ or $\delta_i$ is correctly specified. The third term $\Lambda_{3n,\delta}\left(\alpha_0 \right)$ is the leading bias term. When $\pi_i$ or $\delta_i$ is correctly specified, simple calculations show that 
\begin{align}
\begin{split}
 {\it bias } \left\{ \Lambda_{3n,\delta}\left(\alpha_0 \right)  \right\} &= E \left\{K_h(X-x) V^{-1}\left(x, \boldsymbol{\alpha}_0 \right) \left[ g_0(X) - G (X-x)^T \boldsymbol {\alpha}_0 \right] G(X-x)\right\}\\
 &\qquad\qquad\qquad\qquad\qquad\qquad\qquad\qquad\qquad\qquad\qquad\qquad\qquad+o_p(1)\\
 & = \frac{1}{2}g^{\prime\prime}(x)V^{-1}\{g_0(x)\}f_X(x)H(K)+o\left(h^2 \right)
 \end{split}
\end{align}
where $H(K)$ is a $2 \times 1$ vector with the $k$th element $c_{k+1}(K)\times h^{(k+1)}$. Applying these results to (\ref{eq:discon}), we have the asymptotic bias of the estimator:
\begin{align}
 {\it bias } \left\{ \hat {\alpha}_{DR}(x) \right\} = \frac{1}{2}h^2 g^{\prime\prime}(x)c_2(K)+o\left(h^2 \right)
\end{align}

Now study $\Lambda_{1n,\delta}\left(\alpha_0 \right)-\Lambda_{2n,\delta}\left(\alpha_0 \right)$, which contributes to the leading variance and asymptotic normality.
Note that the variance of $\Lambda_{3n,\delta}\left(\alpha_0 \right)$ is of order $o(1/nh)$, and hence can be ignored asymptotically. Under the ignarability assumption (\ref{eq:ign}), we have $E[D|Y,X,W] = E[D|X,W] = \pi_0(X,W)$, the true conditional mean of $[D|X,W]$. It follows that when either $\pi$ or $\delta$ is correctly specified, $\Lambda_{1n,\delta}\left(\alpha_0 \right)-\Lambda_{2n,\delta}\left(\alpha_0 \right)$ is asymptotically normal with mean 0 and variance
\begin{align}
 {\it var } \left\{ \Lambda _ { 1 n , \delta } \left( \boldsymbol { \alpha } _ { 0 } \right) - \Lambda _ { 2 n , \delta } \left( \boldsymbol { \alpha } _ { 0 } \right) \right\} = \frac { 1 } { n } \left[ {\it var } \left\{ \boldsymbol { \Lambda } _ { 1,2 , \delta } \left( \boldsymbol { \alpha } _ { 0 } \right) \right\} \right]
\end{align}
where
\begin{align}
\begin{split}
\Lambda _ { 1,2 , \delta }& \left( \boldsymbol { \alpha } _ { 0 } \right) = K_h(X-x) V^{-1}\left(x,\boldsymbol{\alpha}_0\right) \boldsymbol{G}(X-x)\\
&\times \left\{  D\left( \frac { (1-Z)D } { \widetilde { \pi } (X,W) } [ Y - g_0(X) ] - \left\{ \frac {(1-Z)D}{ \widetilde {\pi}(X,W)}-1 \right\} \left[\widetilde{\delta}_0 (X,W) - g_0(X) \right] \right)\right.\\ 
&\left. \qquad+ (1-D)\left( \frac { (1-Z)(1-D) } { 1-\widetilde { \pi } (X,W) } [ Y - g_0(X) ] - \left\{ \frac {(1-Z)(1-D)}{ 1-\widetilde {\pi}(X,W)}-1 \right\} \right.\right.\\ 
&\left.\left. \qquad \qquad \qquad \qquad \qquad \qquad \qquad \qquad \qquad \qquad \quad 
\times \left[\widetilde{\delta}_0 (X,W) - g_0(X)\right] \right)
\right\}
\end{split}
\end{align}
Further calculations show that
\begin{align}
\begin{split}
 \frac{1}{n} &\operatorname{var} \left\{\Lambda_{1,2,\delta} (\alpha_0) \right\}\\
 & = \frac{1}{n} E\left[K_h^2 (X - x) V^{-2}( x, \alpha_0) G( X-x )G( X-x )^T\right.  \\
 & \left. \qquad \times \left\{  D\left( \frac { (1-Z)D } { \widetilde { \pi } (X,W) } [ Y - g_0(X) ] - \left\{ \frac {(1-Z)D}{ \widetilde {\pi}(X,W)}-1 \right\} \left[\widetilde{\delta}_0 (X,W) - g_0(X) \right] \right)\right.\right.\\ 
&\left.\left. \qquad+ (1-D)\left( \frac { (1-Z)(1-D) } { 1-\widetilde { \pi } (X,W) } [ Y - g_0(X) ] - \left\{ \frac {(1-Z)(1-D)}{ 1-\widetilde {\pi}(X,W)}-1 \right\} \right.\right.\right.\\ 
&\left.\left.\left. \qquad\qquad\qquad\qquad\qquad\qquad\qquad\qquad\qquad\qquad\qquad 
\times \left[\widetilde{\delta}_0 (X,W) - g_0(X)\right] \right)
\right\}^2\right]  \\
&= \frac{1}{nh} f_X(x) V^{-2} \{ g_0(x) \} E\left[\left\{  D\left( \frac { (1-Z)D } { \widetilde { \pi } (X,W) } [ Y - g_0(X) ] - \left\{ \frac {(1-Z)D}{ \widetilde {\pi}(X,W)}-1 \right\} \right.\right.\right.\\ 
&\left.\left.\left. \qquad\qquad\qquad\qquad\qquad\qquad\qquad\qquad\qquad\qquad\qquad\qquad 
\times \left[\widetilde{\delta}_0 (X,W) - g_0(X) \right] \right)\right.\right.\\ 
&\left.\left. \qquad+ (1-D)\left( \frac { (1-Z)(1-D) } { 1-\widetilde { \pi } (X,W) } [ Y - g_0(X) ] - \left\{ \frac {(1-Z)(1-D)}{ 1-\widetilde {\pi}(X,W)}-1 \right\} \right.\right.\right.\\ 
&\left.\left.\left. \qquad\qquad\qquad\qquad\qquad
\times \left[\widetilde{\delta}_0 (X,W) - g_0(X)\right] \right)
\right\}^2| X = x \right]  D\left(K^2 \right) + o \left(\frac{1}{nh} \right)
\end{split}
\end{align}
Applying these results to (\ref{eq:discon}) and Theorem 2 follows.


\newpage

\begin{thebibliography}{99}

\bibitem{Angrist2015}Angrist, J. D., and Rokkanen, M. (2015), ``Wanna Get Away? Regression Discontinuity Estimation of Exam School Effects Away From the Cutoff," {\it Journal of the American Statistical Association}, 110:512, 1331-1344.

\bibitem{Calonico2014}Calonico, S., Cattaneo, M. D., and Titiunik, R. (2014), “Robust nonparametric bias corrected inference in the regression discontinuity design,” {\it Econometrica}, 82, 2295-2326. 

\bibitem{Carroll1997}Carroll, R. J., Iturria, S. J., and Gutierrez, R. G. (1997), ``Estimating Covariance Matrices Using Estimating Functions in Nonparametric and Semiparametric Regression,” in {\it Selected Proceedings of the Symposium on Estimating Functions}, Hayward, CA: Institute of Mathematical Statistics, 399-404.

\bibitem{Cattaneo2016}Cattaneo, M. D., Keele, L., Titiunik, R., and Vazques-Bare, G., (2016), “Interpreting Regression Discontinuity Designs with Multiple Cutoffs” {\it Journal of Politics}, 78, 1229-1248.

\bibitem{Chiou2018}Chiou, Y. Y., Chen, M. Y., and Chen, J., (2018), “Nonparametric Regression with Multiple Thresholds: Estimation and Inference” {\it Journal of Econometrics}, 206, 472-514.

\bibitem{Crost2014}Crost, B., Felter, J. H., and Johnston, P. B. (2014), “Aid Under Fire: Development Projects and Civil Conflict,” {\it American Economic Review}, 104, 1833-1856.

\bibitem{Fan1996}Fan, J., and Gijbels, I. (1996), {\it Local Polynomial Modelling and Its Applications}, London: Chapman \& Hall.

\bibitem{Hahn2001}Hahn, J., Todd, P.  and Van Der Klaauw, W. (2001), ``Identification and Estimation of Treatment Effects with a Regression-Discontinuity Design,” {\it Econometrica}, 69, 201-209.

\bibitem{}H\"{a}rdle, W. (1990), {\it Applied Nonparametric Regression}, Cambridge University Press.  

\bibitem{}H\"{a}rdle, W., M\"{u}ller, M., Sperlich, S., Werwatz, A. (2004), {\it Nonparametric and Semiparametric Models}, Springer-Verlag Berlin Heidelberg.  

\bibitem{Henderson2014}Henderson, D. J., Parmeter, C. F., and Su, L. (2014), “Nonparametric threshold regression: Estimation and inference” Working paper, Department of Economics, University of Miami.

\bibitem{Hendeson2015}Henderson, D. J., and Parmeter, C. F. (2015), {\it Applied Nonparametric Econometrics}, Cambridge University Press book.  

\bibitem{Holand1986}Holland, P. (1986), ``Statistics and causal inference (with discussion)," {\it Journal of the American Statistical Association}, 81, 945-970.

\bibitem{Hoshino2009}Hoshino, T. (2009), {\it Chousa kannsatsu d$\overline{e}$ta no toukei kagaku} (Statistical sciene of survey observation data), Tokyo: Iwanami Shoten. 

\bibitem{Imai2013}Imai, K., and Ratkovic, M. (2013), ``Estimating Treatment Effect Heterogeneity in Randomized Program Evaluation", {\it The Annals of Applied Statistics}, 7, 443-470.

\bibitem{Imbens2008} Imbens, G., and Lemieux, T. (2008), ``Regression Discontinuity Designs: A Guide to Practice,'' {\it Journal of Econometrics}, 142, 615-635.

\bibitem{Imbens2012}Imbens, G. W. and Kalyanaraman, K. (2012), “Optimal bandwidth Choice for the Regression Discontinuity Estimator,” {\it The Review of Economic Studies}, 79, 933-959. 

\bibitem{Lee2008}Lee, D. S. (2008), “Randomized Experiments from Non-random Selection in U.S. House Elections,” {\it Journal of Econometrics}, 142, 675-697.

\bibitem{Lee2010} Lee, D. S., and Lemieux, T. (2010), ``Regression Discontinuity Design in Economics,” {\it Journal of Economic Literature}, 48, 281-355.

\bibitem{Li2007} Li, Q., and Racine, J. S. (2007), {\it Nonparametric Econometrics: Theory and Practice}, Princeton University Press.

\bibitem{Liang1986}Liang, K. Y., and Zeger, S. L. (1986), ``Longitudinal Data Analysis Using Generalized Linear Models," {\it Biometrika}, 73, 1, 13-22.

\bibitem{Lucas2014}Lucas, A.M. and Mbiti, I. (2014), “Effects of School Quality on Student Achievement: Discontinuity Evidence from Kenya,” {\it American Economic Journal: Applied Economics}, 6(3), 234-263.

\bibitem{Ludwig2007}Ludwig, J., and Miller, D. L. (2007), “Does Head Start Improve Children's Life Changes? Evidence from a Regression Discontinuity Design,” {\it Quarterly Journal of Economics}, 122, 159-208.

\bibitem{Papay2011}Papay, J.P., Willett, J. B., Murnane, R. J. (2011), “Extending the regression-discontinuity approach to multiple assignment variables.” Journal of Econometrics, 161, 203-207.

\bibitem{Porter2015}Porter, J. and Yu, P. (2015), “Regression Discontinuity Designs with Unknown Discontinuity Points; Testing and Estimation” {\it Journal of Econometrics}, 189, 132-147.

\bibitem{Racine2003}Racine, J., and Li, Q. (2003), ``Nonparametric Estimation of Regression Functions with Both Categorical and Continuous Data," {\it Journal of Econometrics}, 119, 99-130.

\bibitem{Robins1994} Robins, J. M., Rotnitzky, A., and Zhao, L. P. (1994), “Estimation of Regression Coefficients When Some Regressors Are Not Always Observed,” {\it Journal of the American Statistical Association}, 89, 846-866.

\bibitem{Rosenbaum1983}Rosenbaum, P. R., and Rubin, D. B. (1983), ``The Central Role of the Propensity Score in Observational Studies for Causal Effects," {\it Biometrika}, 70, 41-55.

\bibitem{Rubin1974}Rubin, D. B. (1974), ``Estimating causal effects of treatments in randomized and non-randomized studies," {\it Journal of Educational Psychology}, 66, 688-701.

\bibitem{Rubin1976}Rubin, D. B. (1976), ``Inference and missing data," {\it Biometrika}, 63, 3, 581-592.

\bibitem{Ruppert1997} Ruppert, D. (1997), “Empirical-Bias Bandwidths for Local Polynomial Nonparametric Regression and Density Estimation,” {\it Journal of the American Statistical Association}, 92, 1049-1062.

\bibitem{1960}Thistlethwaite, D. L., and Campbell, D. T. (1960), ``Regression-Discontinuity Analysis: An Alternativeto the Ex Post Facto Experiment,” {\it Journal of Educational Psychology}, 51(6): 309-317.

\bibitem{Wang2010} Wang, L., Rotnitzky, A. and Lin, X. (2010), “Nonparametric Regression with Missing Outcomes Using Weighted Kernel Estimating Equations,” {\it Journal of the American Statistical Association}, 105:491, 1135-1146.


\end{thebibliography}
\end{document}